\documentclass[twocolumn,times,twocolappendix]{aastex631}

\newcommand{\bigO}[1]{$\mathcal{O}$(#1)}
\usepackage{tabularx}
\usepackage{gensymb}
\usepackage{xcolor}
\usepackage{upgreek}
\newcommand{\redtwo}[1]{#1}
\newcommand{\red}[1]{#1}

\begin{document}

\title{Low-Energy Electron-Track Imaging for a Liquid Argon Time-Projection-Chamber Telescope Concept using Probabilistic Deep Learning}
\shorttitle{E-Track Imaging in GammaTPC with Deep Learning}

\correspondingauthor{M. Buuck}
\email{mbuuck@slac.stanford.edu}

\author{M. Buuck}
\affiliation{SLAC National Accelerator Laboratory, Menlo Park, CA 94025, USA}
\affiliation{Kavli Institute for Particle Astrophysics and Cosmology, Stanford University, Stanford, CA 94305, USA}

\author{A. Mishra}
\affiliation{SLAC National Accelerator Laboratory, Menlo Park, CA 94025, USA}

\author{E. Charles}
\affiliation{SLAC National Accelerator Laboratory, Menlo Park, CA 94025, USA}
\affiliation{Kavli Institute for Particle Astrophysics and Cosmology, Stanford University, Stanford, CA 94305, USA}

\author{N. Di Lalla}
\affiliation{Kavli Institute for Particle Astrophysics and Cosmology, Stanford University, Stanford, CA 94305, USA}
\affiliation{Physics Department, Stanford University, Stanford, CA 94305, USA}
\affiliation{Hansen Experimental Physics Laboratory, Stanford, CA 94305, USA}

\author{O.A. Hitchcock}
\affiliation{Physics Department, Stanford University, Stanford, CA 94305, USA}

\author{M.E. Monzani}
\affiliation{SLAC National Accelerator Laboratory, Menlo Park, CA 94025, USA}
\affiliation{Kavli Institute for Particle Astrophysics and Cosmology, Stanford University, Stanford, CA 94305, USA}
\affiliation{Vatican Observatory, Castel Gandolfo, V-00120, Vatican City State}

\author{N. Omodei}
\affiliation{Kavli Institute for Particle Astrophysics and Cosmology, Stanford University, Stanford, CA 94305, USA}
\affiliation{Physics Department, Stanford University, Stanford, CA 94305, USA}
\affiliation{Hansen Experimental Physics Laboratory, Stanford, CA 94305, USA}

\author{T. Shutt}
\affiliation{SLAC National Accelerator Laboratory, Menlo Park, CA 94025, USA}
\affiliation{Kavli Institute for Particle Astrophysics and Cosmology, Stanford University, Stanford, CA 94305, USA}



\begin{abstract}

The GammaTPC is an MeV-scale single-phase liquid argon time-projection-chamber gamma-ray telescope concept with a novel dual-scale pixel-based charge-readout system.  It promises to enable a significant improvement in sensitivity to MeV-scale gamma-rays over previous telescopes. The novel pixel-based charge readout allows for imaging of the tracks of electrons scattered by Compton interactions of incident gamma-rays. The two primary contributors to the accuracy of a Compton telescope in reconstructing an incident gamma-ray's original direction are its energy and position resolution. In this work, we focus on using deep learning to optimize the reconstruction of the initial position and direction of electrons scattered in Compton interactions, including using probabilistic models to estimate predictive uncertainty. We show that the deep learning models are able to predict locations of Compton scatters of MeV-scale gamma-rays from simulated \redtwo{500 $\upmu$m} pixel-based data to better than \redtwo{1} mm root-mean-squared error, and are sensitive to the initial direction of the scattered electron. We compare and contrast different deep learning uncertainty estimation algorithms for reconstruction applications. Additionally, we show that event-by-event estimates of the uncertainty of the locations of the Compton scatters can be used to select those events that were reconstructed most accurately, leading to improvement in locating the origin of gamma-ray sources on the sky.

\end{abstract}



\section{Introduction}
\begin{figure*}
    \centering
    \includegraphics[width=\textwidth]{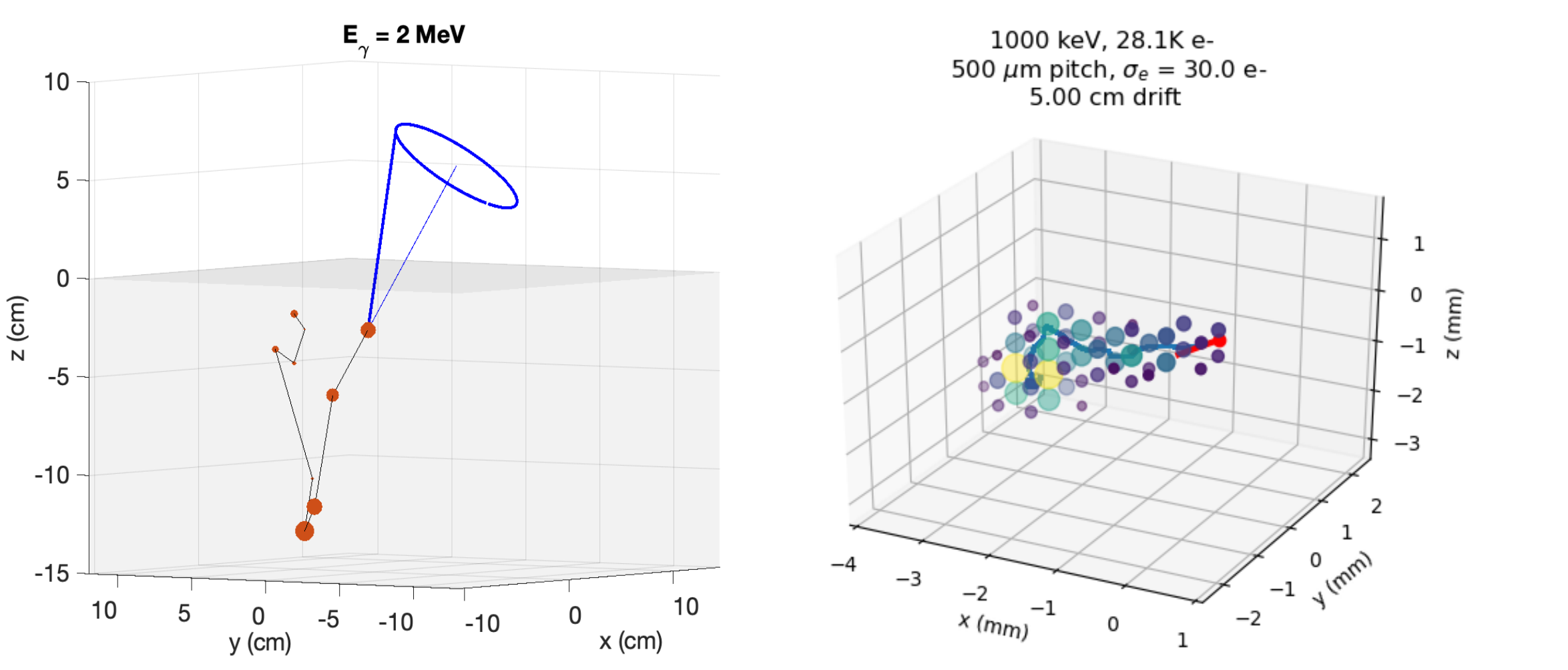}
    \caption{(left) Simulated typical gamma-ray event \red{in a large uniform volume of liquid argon} with multiple scatters, with the reconstructed event ring and the initial (true) gamma ray direction shown, along with a schematic electron recoil track (not to scale). The size of the dots at each scatter is proportional to the deposited energy. (right) Typical 1\,MeV electron recoil track, consisting of (blue line) \redtwo{28,100}\,e$^-$-ion pairs, and the simulated readout of these electrons (circles) from a 500\,$\upmu$m pitch pixel readout with \redtwo{30}\,e$^-$ noise after \redtwo{5}\,cm drift.  The head (red circle) and true initial recoil direction (faint red line) are indicated. The tail of the track has higher charge density (yellow), a result of the energy dependence of $dE/dx$. The gamma-ray scatters (left panel) were simulated with a custom Monte-Carlo-based package, and the electron track (right panel) was simulated with PENELOPE. \citep{penelope}}
    \label{fig:event}
\end{figure*}

Astrophysical sources that produce MeV-scale gamma-rays are some of the least well characterized over the entire electromagnetic spectrum (see, for instance \cite{McEnery:2019tcm}). Such photons Compton scatter multiple times over \red{a relatively large volume (Figure~\ref{fig:event}), and imaging with Compton telescopes requires an accurate measurement of the energy and locations of these scatters with minimal loss of energy in inert material.  This can be a formidable instrumentation challenge, but if accomplished, enables kinematic reconstruction of the path, and hence initial direction, of the gamma-ray as we describe in more detail in Section \ref{sec:gammatpc_inst_concept}}.

\red{We are developing a detector concept, GammaTPC, that is based on liquid argon (LAr) time-projection-chamber (TPC) technology \citep{aramaki2022snowmass2021}. The energy and spatial resolution promises to be comparable, if not better, than that of the central silicon tracking detectors proposed for missions such as AMEGO-X \citep{Fleischhack_2021}.  That is, energy depositions measured at a few percent level near 1 MeV, and, as discussed in this paper, the location of scatter interactions measured to better than $1$ mm, and a powerful determination of the electron track direction.  Moreover, the TPC readout potentially allows this capability over a much larger instrument for the same cost, with a highly uniform response over the active volume.} The science opportunities for such an instrument would be substantial. The most complete catalogs of MeV sources include only dozens of sources \red{\citep{Schoenfelder_2000}}, as compared to the thousands of sources seen by Fermi-LAT~\citep{2020ApJS..247...33A}, or the billions of sources seen or expected by optical surveys such as the Dark Energy Survey~\citep{2019yCat.2357....0A}, Kilo-Degree Survey~\citep{2019A&A...625A...2K}, and Vera Rubin Observatory/Legacy Survey of Space and Time~\citep{2019ApJ...873..111I}. Extrapolations from X-ray and higher-energy gamma-ray data suggest that many source types are expected to have spectral energy density peaks in the MeV range. \redtwo{An instrument with a very large effective area, large field of view, and high efficiency --- which is the goal of our design --- will} be key in observing a large number of transients, including the electromagnetic counterparts to gravitational-wave detections of compact-object mergers, such as GW170817~\citep{2017ApJ...848L..12A}. It will also open a new window to search for signatures of decaying dark matter \citep{2017JCAP...05..001B}. Furthermore, given the presence of nuclear transition lines, as well as the 511 keV electron-positron annihilation line, the bright diffuse galactic signal observed in the MeV band \red{(see, e.g.,~\cite{Weidenspointner_2000})} can provide otherwise unobtainable information about the evolution and particle content of our galaxy.

A key and novel feature of the GammaTPC concept is a high-fidelity pixel-based readout of the electron-recoil tracks that result from each Compton scatter of the gamma-ray, as shown in the right panel of Figure \ref{fig:event}. To make optimal use of this feature, we need methods to accurately determine both the interaction locations (the heads of the tracks) and their initial directions from the pixel-based data. \red{This is possible at these electron energies because the large amount of energy deposited in the Bragg peak at the end of the track makes the tail distinguishable from the head.} Electron track shapes vary widely due to the random nature of the multiple scattering that electrons undergo, \red{which makes algorithmic reconstruction of the initial scattering position and direction nontrivial. \cite{10.1117/12.459381} and \cite{BLACK2007755} used moment-based methods to do this, which were effective for detecting the distribution of polarized X-rays. \cite{LI201762} developed a graph-based technique and showed that it significantly outperformed the moment-based techniques, and \cite{YONEDA2018269} applied a modified version of it to a proposed Compton camera detector.}

\red{Since the development of the moment-based techniques, the field of deep learning has advanced dramatically, most importantly for this case in the area of computer vision. Because a human can usually identify the head and tail of an electron track just from visual inspection based on the Bragg peak (e.g.~in Figure \ref{fig:event}), modern deep learning computer vision techniques \redtwo{have the potential} to perform well in locating the initial location and direction of a Compton-scattered electron. Indeed, \cite{10.1093/ptep/ptab091} have already developed a model based on Convolutional Neural Networks (CNNs) for the Electron Tracking Compton Camera (ETCC), while \cite{PEIRSON2021164740} have developed a model for the IXPE detector making use of \redtwo{ensembles of deterministic neural networks} for uncertainty estimation of the electron track direction.}

\red{In contrast to the ETCC and IXPE, GammaTPC would be a liquid-based detector, and therefore best able to reconstruct electrons with energies of \bigO{100 keV} or above, while the techniques referenced above were all designed for electrons of \bigO{10 keV}. Furthermore, our detector would provide full 3D images of electron tracks, unlike strip-based detectors which create two 2D images to represent 3D information\redtwo{, or pixel detectors like IXPE that integrate the charge over each pixel for the entire event, also producing a 2D image}. Therefore, while some of the same basic concepts that informed \cite{10.1093/ptep/ptab091} also underpin our approach, our models are designed differently to accommodate these distinctions.} We show that we are able to achieve good accuracy in electron track head and initial direction reconstruction with two deterministic 3D CNNs, with mean absolute error in position less than \redtwo{0.8} mm, and information on initial track direction for electron energies as low as 300 keV.

Additionally, having an estimate of the uncertainty in the position and initial direction of Compton-scattered electrons allows us to select data for which the predictive uncertainty is low, which we show improves accuracy in determining the direction of incident gamma-rays. We also expect that reconstruction of the correct sequence of gamma-ray interactions will improve with an uncertainty estimate, \redtwo{although we have not yet investigated this and intend to do so in future work}.

Commonly used deterministic deep learning models do not produce prediction intervals or uncertainties for their predictions, making it impossible to tell when a prediction may be reliable. As an example, state of the art deterministic neural networks are frequently unable to recognize out of sample examples and habitually make incorrect predictions for such cases \citep{amodei2016concrete,nguyen2015deep, hendrycks2016baseline}. To this end, we compare and contrast different deep learning uncertainty quantification (UQ) approaches for predicting the initial location and direction of Compton-scattered electrons. We select the best performing UQ method and apply it to the reconstruction problem. We hope this will provide a guide-map for subsequent investigators, for instance in similar reconstruction tasks involving particle tracking data.  

\section{The GammaTPC Instrument Concept and Compton Event Reconstruction}
\label{sec:gammatpc_inst_concept}

Liquid-noble TPC technology has undergone tremendous development in recent years, having a transformative impact in direct dark matter searches ~\citep{Aprile_2018, Akerib_2020}. It is now the chosen technology for the massive DUNE neutrino program ~\citep{abi2020deep}, showing the scale of investment the particle physics community is making into this technology. In an upcoming article ~\citep{gammatpcpaper}, we will discuss in detail how this technology will be implemented for GammaTPC. \redtwo{Here we provide a description of how the parts of the instrument relevant to the imaging of electron tracks will function.}

\begin{figure*}
    \centering
    \includegraphics[width=\textwidth]{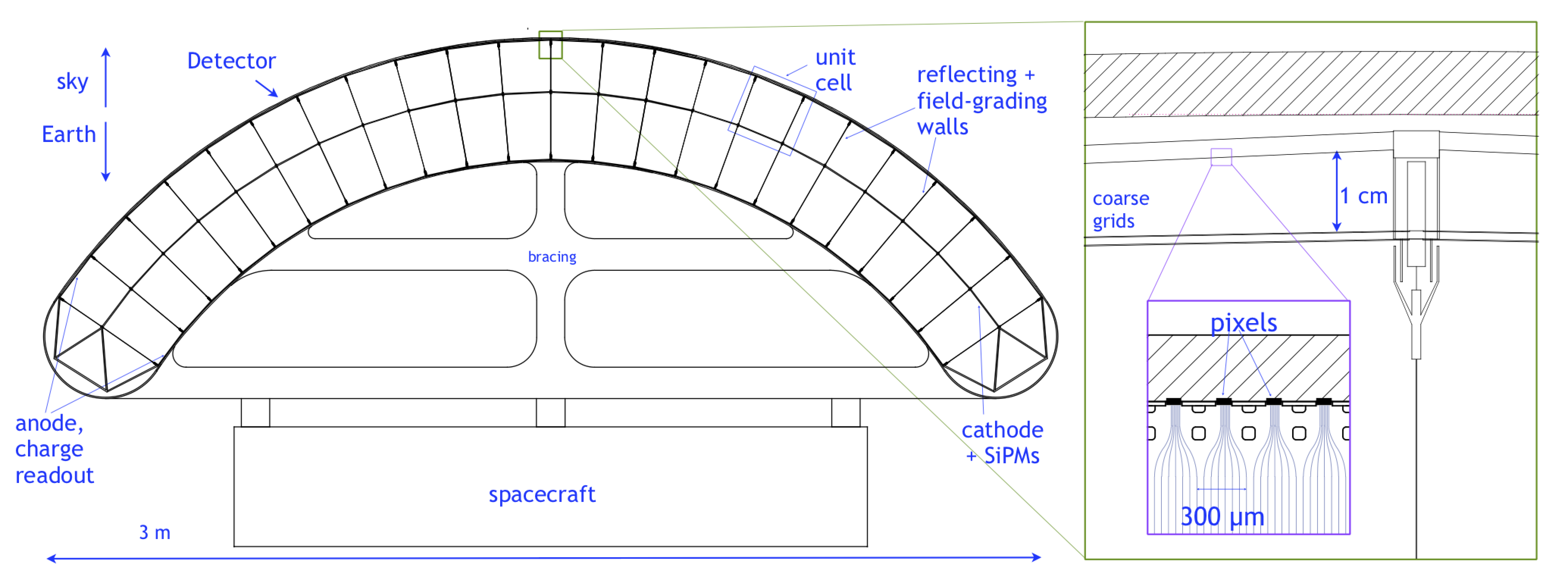}
    \caption{Cross-section schematic of a 10\,m$^2$,  4 tonne implementation of the GammaTPC concept, along with details of the dual-scale charge readout system.  A 40\,cm thick layer of LAr is contained in a thin-walled carbon-fiber shell, along with a lightweight interior cellular readout structure. The overall geometry is a spherical cap.}
    \label{fig:gammatpc_schematic}
\end{figure*}

The GammaTPC instrument concept is shown schematically in Figure \ref{fig:gammatpc_schematic}. Particle interactions in the liquid argon target create scintillation light and free electrons.  An applied electric field drifts electrons to the anode readout planes where their \redtwo{collection on the pixels is digitized, enabling measurement of their} $X-Y$ locations, effectively pixelating the charge readout \redtwo{in 3 dimensions}. \redtwo{Measurement of the light by silicon photomultipliers on the cathode planes establishes} the depth of events, $Z$, as the time difference between the \red{prompt scintillation signal (with 6 ns and 1.6 $\upmu$s time constants)} and the arrival of the slower ($\sim$170\,$\upmu$s over 20\,cm) drifting electrons.  The event energy is inferred from a combination of the charge and scintillation signals.  The high particle rate in low earth orbit combined with the relatively slow charge drift requires vertical segmentation of approximately 20 cm.  The curved geometry provides a very large field of view with relatively uniform response, and also minimizes the mass of the carbon-fiber pressure vessel.  A 10 kV voltage difference between anode and cathode establishes a 0.5 kV/cm charge drift-field. The lateral segmentation is provided by thin reflecting walls.  The overall detector thickness, here 40\,cm, is chosen to have a high interaction efficiency for gamma-rays up to $\sim$10\,MeV.

The core advantage of a TPC is that it provides 3D readout of a uniform target volume with sensors deployed on only a portion of the 2D surfaces.  This enables a large instrument with relatively few channels, and hence low cost and power. This in turn allows a high-granularity readout, which directly leads to good angular resolution, while the minimal interior dead material maximizes event reconstruction efficiency. In our case, the readout, which builds on recent advances in cryogenic complementary metal-oxide semiconductor (CMOS) charge-readout ~\citep{CryoAsic1, Dwyer_2018, Adams_2020}, uses a triggered, ultra low power sub-mm pixel readout to provide sub-mm 3D sampling of the tracks. \redtwo{The trigger for pixel readout is provided by a separate set of ``coarse grid" wire electrodes with a nominal $\sim$1 cm pitch.  This wide spacing avoids loss of signal due to diffusion spreading the signal over several channels and thus will provide the charge integral.  When read out with cryogenic CMOS technology, we expect noise of $\sim$20 electrons per wire of 2-3 pF capacitance \citep{Deng_2018},  with the signal shared amongst several wires. Initial studies show an energy threshold per scatter set by this coarse grid readout near 10-20 keV.  The event energy comes from the summed coarse grid charge signals from all scatters, combined with the integral light signal. The pixel readout, by contrast, is used primarily to image the electron tracks for extraction of the location and direction of Compton electron recoils, the analysis of which is the focus of this paper.  The pixels have substantially lower capacitance than the coarse wires, but also shorter shaping times, and we estimate their noise to be 20-30 electrons.}

A key event reconstruction challenge for this technology is to correctly determine the true sequence of the interactions, or at minimum, correctly establish the first two interactions.  This is primarily done by kinematic testing of possible sequences, with the precision of the measurements of the energies and positions of the interactions driving the efficiency of selecting the correct sequence. The direction of the original gamma-ray then lies at an angle relative to the vector established by the locations of the first two Compton scatters, as seen in Figure \ref{fig:event}. This angle is determined by the energy lost by the gamma-ray in the first Compton interaction. This creates a ring on the sky of possible locations for the original gamma-ray. For an ensemble of gamma-rays from a given source, the overlap of these rings gives the specific location of the source. Because neither the energies nor interaction locations are perfectly measured, there is an effective width to these rings, and therefore an uncertainty in the reconstructed source position. Minimizing this width, i.e.~achieving the best possible pointing, is a key driver of the overall instrument sensitivity, and is achieved by having the most precise and accurate measurements of the energies and locations of the Compton interactions, which also maximizes the efficiency and power of the kinematic tests used to reconstruct the event. \red{It is also important to choose a low Z material to minimize Doppler broadening~\citep{Zoglauer_2003}.}

If the initial track direction can also be estimated, the ring reduces to an arc on the sky, depending on the precision of the measurement. \red{We also expect that the initial track direction estimate will improve the quality of kinematic tests used for reconstruction, though the power of this has not yet been quantified. This would have the strong benefit of reducing the number of gamma-rays needed to locate a source.}

\section{Machine Learning Based Electron Track Reconstruction}
\label{subsec:data_sims}

We now turn to the application of data driven modeling techniques to electron track reconstruction and the estimation of predictive uncertainty, which is the primary focus of this work. There are a variety of different modeling algorithms that can be used for the electron track reconstruction task. A guiding \redtwo{metric for selection} amongst them would be \redtwo{the degree to which} the inductive bias of each algorithm agrees with the nature of the problem. Inductive bias refers to the set of assumptions a learning algorithm uses to generalize beyond its training data \citep{mitchell1980need, baxter2000model}. For the estimation of the track head and initial direction, we require translational invariance. In deep learning architectures, convolutional layers are equivariant to spatial translation. When coupled with pooling layers, they are approximately translation invariant. Consequently, CNN-based feature extraction is not affected by the absolute position of the feature in the feature map. Thus, we use a CNN-based model for estimation, treating the pixelated electron track as a 3D image. We use this model to estimate the track head, the initial scattering direction of each electron, as well as the uncertainty on each estimate produced by the model.\footnote{\red{Jupyter notebooks that will reproduce the results in this paper can be found at \url{https://gitlab.com/probabilistic-uncertainty-for-gammatpc}. Other code used in this study, such as the detector response model, can be made available upon request.}}

\begin{figure*}
    \centering
    \includegraphics[angle=90,height=0.9\textheight]{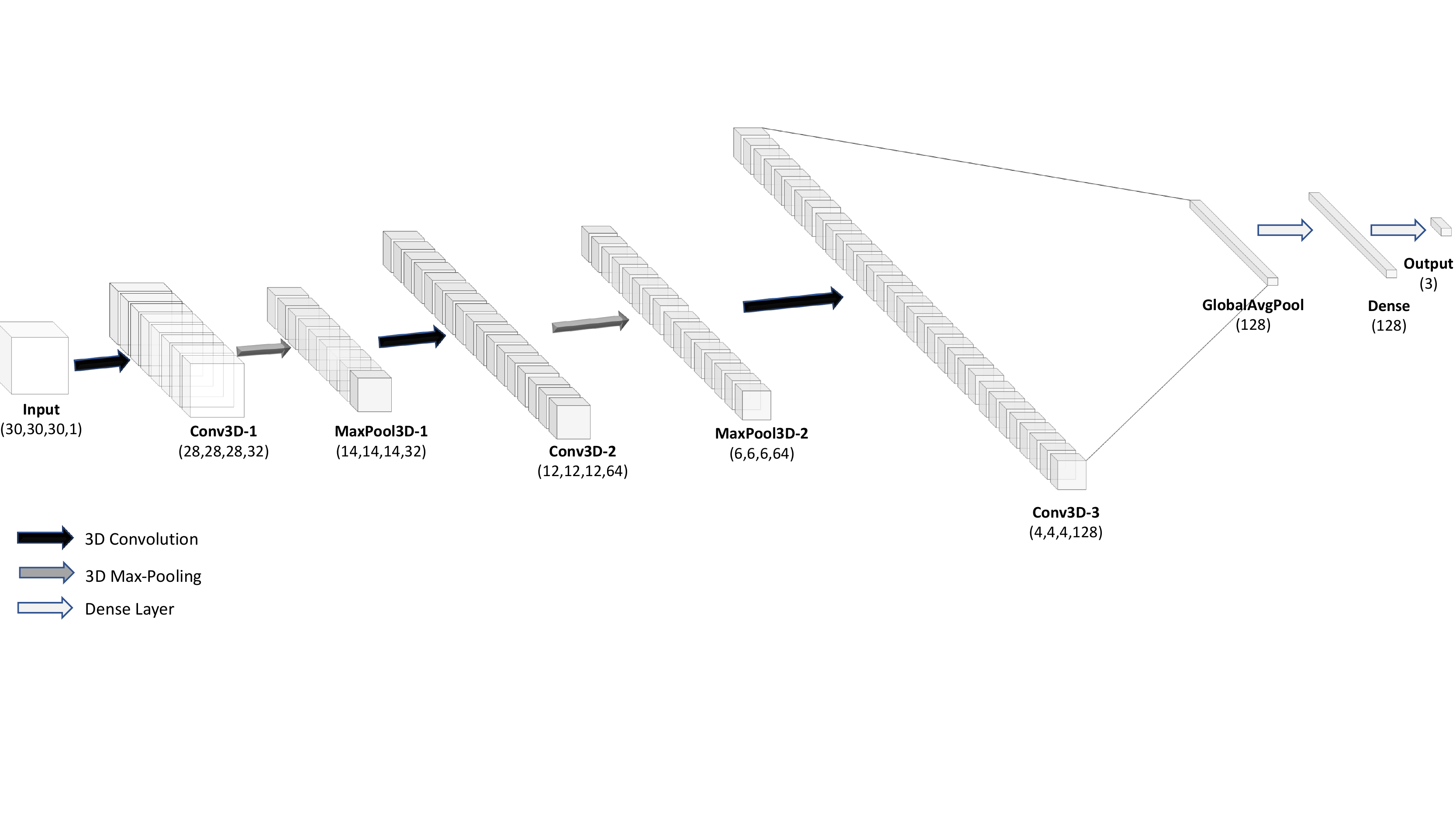}
    \caption{Schematic outlining the architecture of the convolutional neural network. The shapes of the layer outputs are reported, along with representative cognomen for individual layers.}
    \label{fig:3DCNNDiagram}
\end{figure*}

\subsection{Data simulation and model evaluation technique}
\label{subsec:data_sim}
We prepared the training data used to develop the models presented in this paper in a two-step process. We used the PENetration and Energy LOss of Positrons and Electrons (PENELOPE) code \citep{penelope} to generate raw electron tracks, and then subsequently applied detector-response effects with a custom code. \redtwo{PENELOPE provides a detailed microphysical simulation of electron energy loss processes in matter, including simulations of elastic and inelastic scattering off of other particles in the media, bremsstrahlung emission, and generation of delta-rays.} The custom code voxelizes the raw-track energy depositions, simulating digitized pixel readout with a specified pixel pitch determining the $X-Y$ pixel spacing. \red{The $Z$ spacing is determined by the specified electric field, electron mobility, and digitization frequency. The electric field is set to 500 V/cm, and the electron mobility is calculated according to Equation 21 from \cite{LI2016160}. The digitization frequency is then set to achieve an effective $Z$ spacing equal to the $X-Y$ spacing.} For the track head reconstruction, we simulated data with a \redtwo{500 $\upmu$m pixel pitch at \redtwo{4} different drift depths (1 cm, 5 cm, 10 cm, \redtwo{and 20 cm})\footnote{\redtwo{These parameters are a good starting point based on initial instrument design considerations, but may eventually change as the design evolves.}}.} The training data for the track direction reconstruction also uses a 500 $\upmu$m pitch, and is simulated at a drift depth of 5 cm. Each sample therefore consists of a 3D array of floating-point numbers, representing a simulated pixel readout at each position in a 3D grid. The size of this grid in each training data set is given by the bounding box of the largest sample in that data set. All other simulated charge depositions are zero-padded symmetrically to match that dimension.

All simulations compute detected quanta from first principles based on raw energy depositions in a detailed manner similar to that of \redtwo{the Noble Element Simulation Technique (NEST)} \citep{Szydagis_2011}. All simulations also have a transverse diffusion coefficient of 12 cm${}^2$/s, taken from \cite{LI2016160}. The longitudinal diffusion coefficient is computed from Equation 23 of \cite{LI2016160}, and is approximately 6 cm${}^2$/s, giving $\sim$0.4 mm root-mean-squared (RMS) diffusion over a 10 cm drift. We also apply an effective electronic noise of approximately \redtwo{30} electrons, and a simulated digitization threshold of 3$\sigma$ above noise. We simulated a  number of electrons equal to within 6\% at 5 energies (50, 300, 500, 750, and 1,000 keV), obtaining 24,472 simulated electrons in total. We then split the data randomly into training and testing sets in a 90:10 ratio for the track head and direction deterministic models, and a 92.5:7.5 ratio for the track head uncertainty quantification model.

We assess subsequent performance of the trained models not only with standard measures such as accuracy and coverage, but also with figures-of-merit for the quality of the overall Compton event reconstruction. We do this using data simulated with the Medium-Energy Gamma-ray Astronomy library, or MEGAlib \citep{MEGALIB}. After implementing the detector geometry using MEGAlib's built-in geometry specification, we generate \redtwo{1,000 keV gamma-rays originating at 64 degrees from the zenith} using the MEGAlib \texttt{FarFieldPointSource} generator. We then process the generated energy depositions with custom code that applies the detector response, including a parameterization of the interaction locations (with uncertainties) based on the output of the trained neural networks on test electron track simulations. We present a description of this parameterization in Sections \ref{sec:det_param} and \ref{sec:edl_param}. We then feed this simulated detector output, in the form of a MEGAlib-specific file type, into the MEGAlib reconstruction code \texttt{revan}, which determines the time-ordering of the Compton scatters. At this point we \redtwo{remove any events where \texttt{revan} was unable} to determine the first two Compton scatters correctly. \redtwo{Because events with an incorrect reconstruction of the scatter order tend to have significantly larger pointing error, this selection will produce a smaller pointing error than would occur in a real experiment. However, since we are primarily concerned with understanding specifically the behavior of the electron scatter position reconstruction, and not the Compton ordering reconstruction, we choose to do this so as to eliminate that possibly confounding source of error from our assessment of the performance of our models.} These events are then then used by \red{a modified version of} the \href{https://github.com/ComPair/python}{ComPair Python library} \citep{compair} to compute the location of and uncertainty in the gamma-ray origin.

\subsection{Electron Track Head Reconstruction}
\label{subsec:deterministic_model}

\redtwo{In the first model,} we use a deterministic CNN-based model to predict the location of the electron track head. We report the architecture of the CNN used for electron track head reconstruction in Figure \ref{fig:3DCNNDiagram}. The input to the model is a 3D image of an electron track from the detector. We include spatial dropout \citep{tompson2015efficient} (as opposed to conventional dropout \citep{dropout}) and Batch Normalization layers after each convolutional layer to ameliorate overfitting. We use a 3-dimensional global average pooling operation at the end of the feature extraction stage, as opposed to a conventional flattening operation, to reduce the parameters in the ensuing fully connected layers and improve generalization. The final output from the fully connected layers is a 3-dimensional regression prediction for the location of the electron track head. The model architecture and ancillary hyperparameters were established based on Bayesian Optimization \citep{omalley2019kerastuner}, followed by subsequent manual fine tuning. \red{This model was implemented with Keras \citep{chollet2015keras}, using the Adam optimizer \citep{adam_optimizer}, a mean-squared-error loss function, and was trained for 200 epochs on an NVIDIA Tesla A100 GPU at the SLAC Scientific Data Facility, which took about 10 minutes.}

\begin{figure}
    \centering
    \includegraphics[width=\columnwidth, trim={0.6cm 0.7cm 0.6cm 0.65cm},clip]{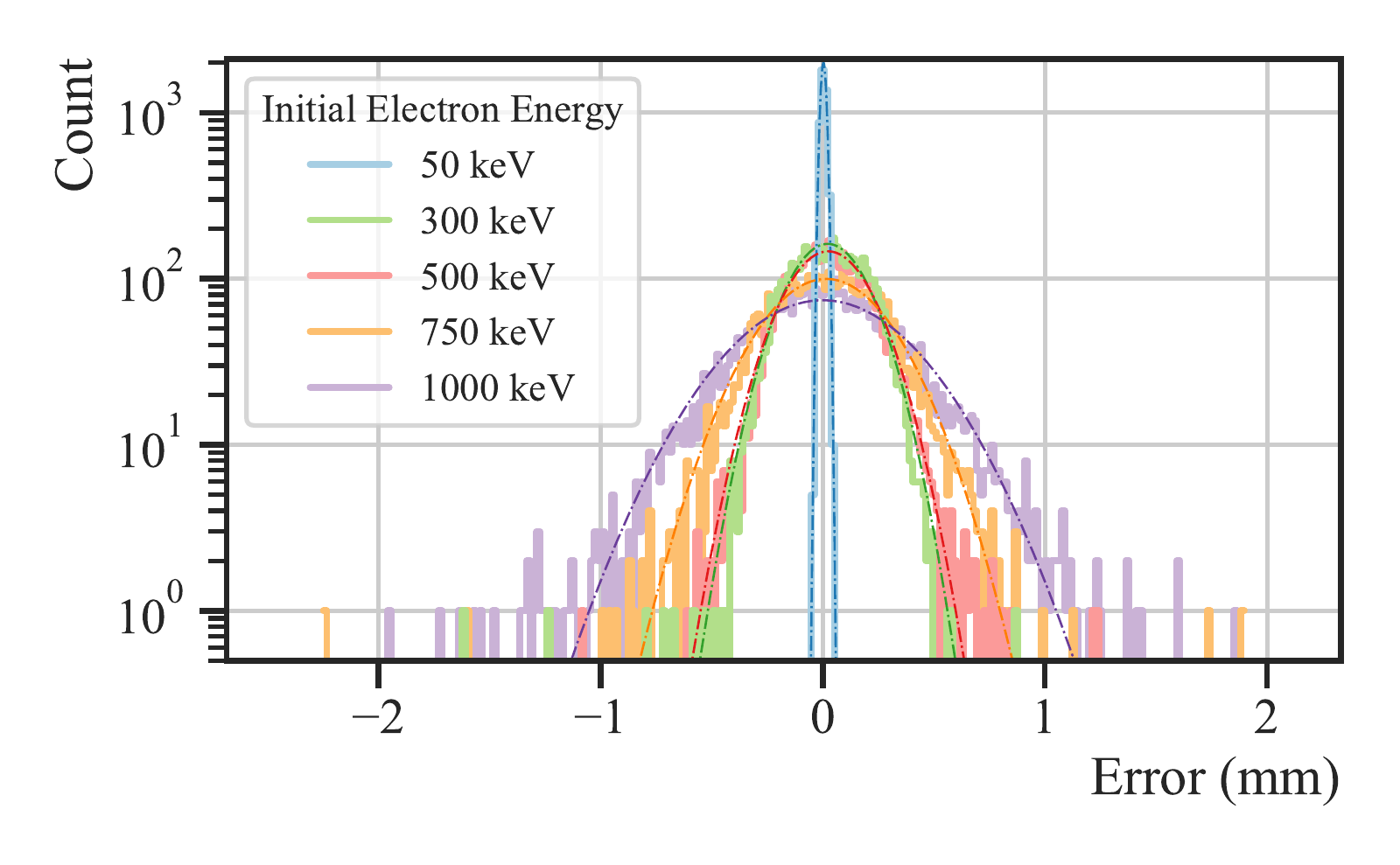}
    \caption{Empirical error distribution of the trained deterministic model for a 5 cm drift depth, broken down by initial electron energy and fitted to Gaussian distributions. This \redtwo{figure contains all of the $X$, $Y$, and $Z$ errors, and does not sum them in quadrature before filling the histogram.}}
    \label{fig:fitted_error}
\end{figure}

\begin{figure}
    \centering
    \includegraphics[width=\columnwidth]{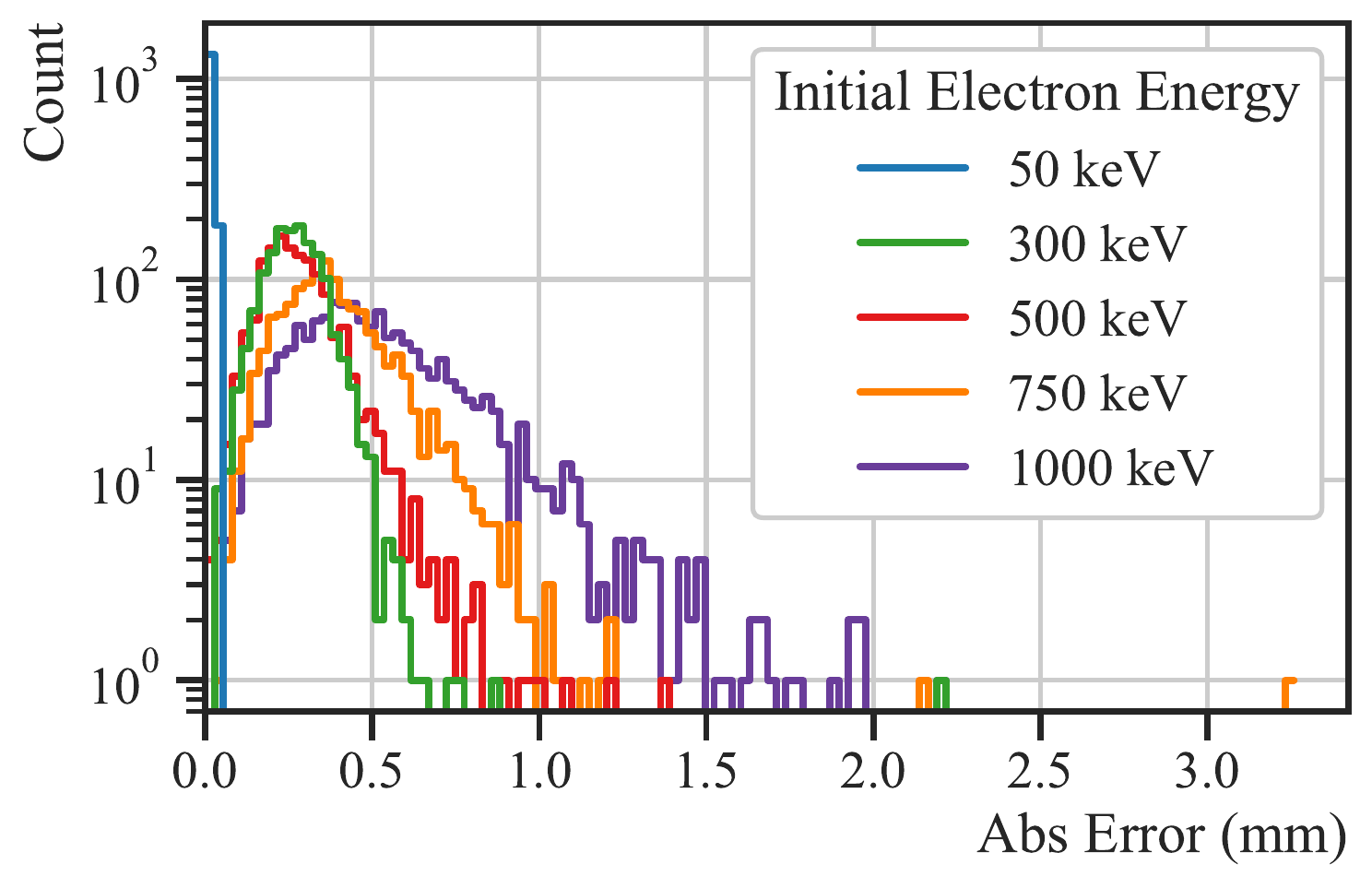}
    \caption{Empirical error distribution of the trained deterministic CNN. This \redtwo{figure gives the actual distance} between the true and predicted electron track head location.}
    \label{fig:deterministic_5_cm_drift_abs_error}
\end{figure}

Since the model predicts the location of the $X$, $Y$, and $Z$ components of the electron track head independently, and treats each physical dimension identically, we can combine the predictions of all 3 dimensions and analyze them in the same distribution. The distribution of errors on the test data set in all 3 dimensions is shown in Figure \ref{fig:fitted_error}, broken down by initial electron energy.  However, from an applications perspective, the goal is a sub-mm accuracy of the electron track head location across all three spatial dimensions. Therefore, we add in quadrature the $X$, $Y$, and $Z$ error for each prediction to obtain the absolute error (i.e.~the Euclidean distance between the true and predicted electron track head position) for each sample in the test data set. We show this distribution in Figure \ref{fig:deterministic_5_cm_drift_abs_error}. Clearly the distribution of errors is wider for higher energy electron tracks. \redtwo{Because the beginning of an electron track tends to become straighter as the electron's initial energy increases (this effect is illustrated in more detail in Section \ref{subsec:direction_model}),} one might expect that the beginning of the track should therefore become \textit{easier} to identify, not harder. We hope to investigate this with a future interpretability study.

\begin{figure}
    \centering
    \includegraphics[width=\columnwidth, trim={0.7cm 0.7cm 0.6cm 0.65cm},clip]{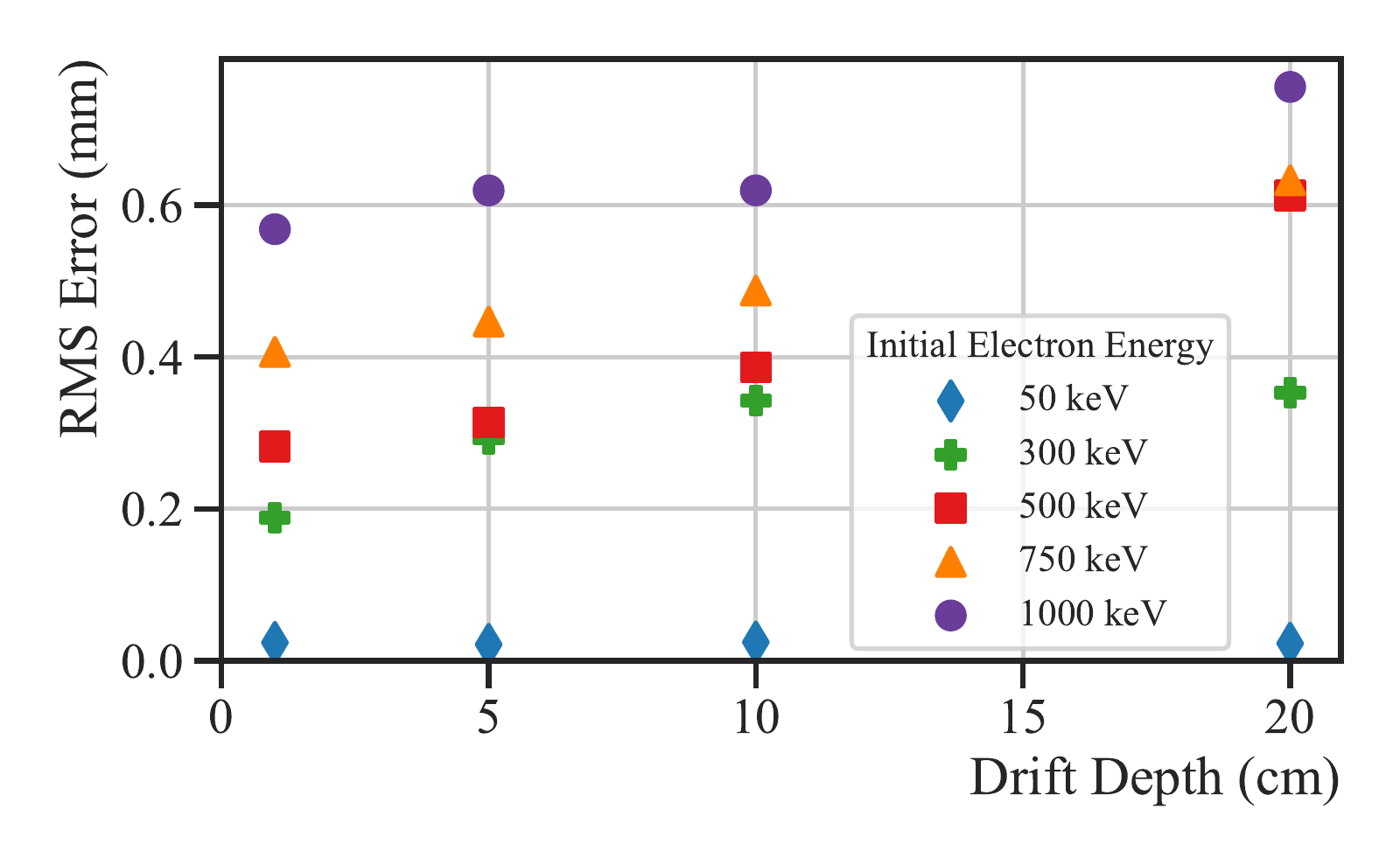}
    \caption{The \redtwo{square root of the mean of the distribution of the squared differences between the predicted and true track head location (as in Figure \ref{fig:deterministic_5_cm_drift_abs_error} with squared errors)} for different initial electron energies as a function of drift depth.}
    \label{fig:det_fwhm_vs_drift}
\end{figure}

We plot the mean of each of these distributions in Figure \ref{fig:det_fwhm_vs_drift}, as a function of drift depth. As expected, the RMS error increases for all energies with drift depth, consistent with diffusion effects spreading out the signal and making it harder to pinpoint the track head location. However it is important to note that this increase with drift length is modest, and we readily achieve sub-mm accuracy up to at least a \redtwo{20} cm drift length. There are no error bars given for the points in Figure \ref{fig:det_fwhm_vs_drift} because the statistical errors are negligible, and the deterministic approach does not provide a good way of estimating a systematic uncertainty. We approach this problem with uncertainty estimation techniques in Section \ref{subsec:edl_model}.

\begin{figure}
    \centering
    \includegraphics[width=\columnwidth, trim={0.7cm 0.7cm 0.6cm 0.65cm},clip]{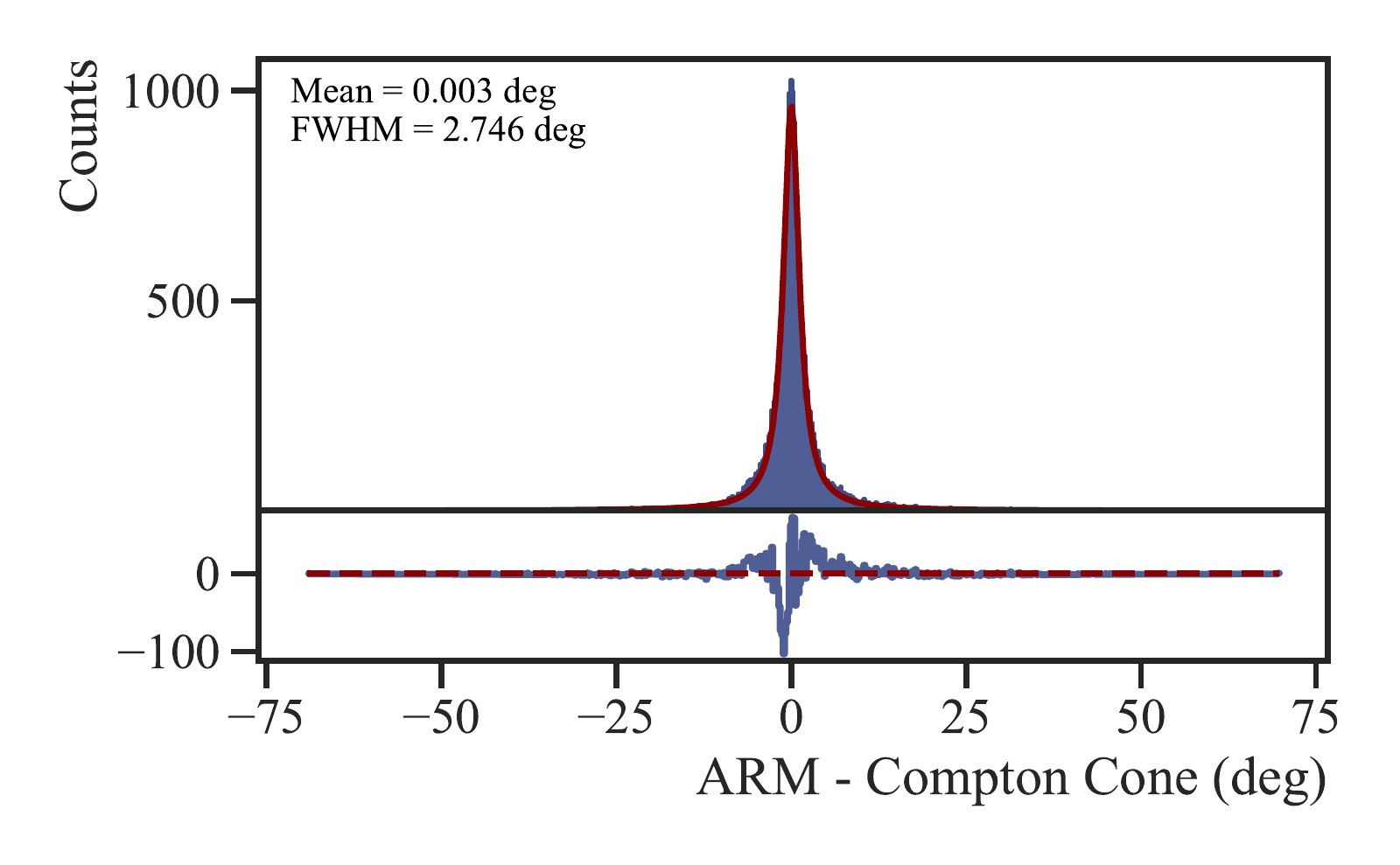}
    \caption{Histogram of the difference in degrees between the true origin of the gamma-ray and the closest point on the reconstructed Compton ring. It is fitted to a Lorentz distribution.}
    \label{fig:det_ARM}
\end{figure}

In Figure \ref{fig:det_ARM} we present an assessment of the deterministic model using the MEGAlib-simulated performance of the GammaTPC instrument. \redtwo{We apply a parameterization of the deterministic model output to the simulated data produced by MEGAlib and our custom detector-effects code, which is described in detail in Appendix Section \ref{sec:det_param}.} The distribution of the Angular Resolution Measure (ARM), which is the angular difference between the true and reconstructed origin of gamma-rays, is peaked at zero degrees, and is shown fitted to a Lorentz distribution\footnote{\red{We also attempted a fit to a Voigt profile, but the best fit contained no Gaussian component and was therefore also a Lorentzian.}}. The Full-Width-at-Half-Maximum (FWHM) of this distribution for a 5 cm drift depth (as shown) is \redtwo{2.746\degree}.

\vspace{50mm}
\subsection{Electron Scatter Direction Reconstruction}
\label{subsec:direction_model}
We also developed a model to reconstruct the direction of the initial electron scatter. Obtaining the direction is very useful, because, as explained in Section \ref{sec:gammatpc_inst_concept}, breaking the axial degeneracy of the reconstructed gamma-ray origin can greatly improve the pointing accuracy of a well-reconstructed event.

\begin{figure}
    \centering
    \includegraphics[width=\columnwidth]{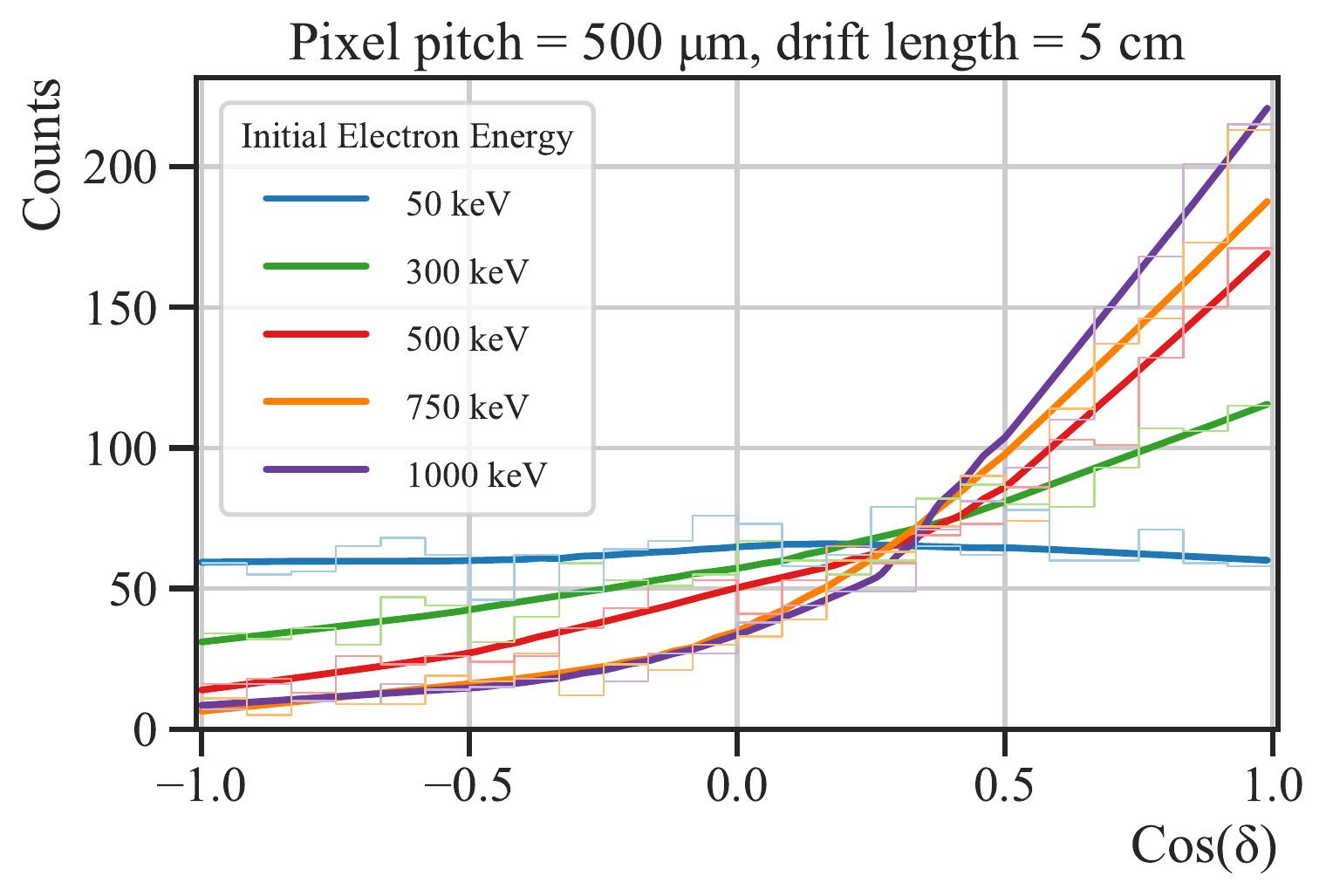}    
    \caption{Empirical distribution of the cosines of the angle between the reconstructed and true direction of initial electron scatter. The light-colored histogram contains the predictions from the test data set, and the dark lines are local regressions on the histogram count values. The local regressions are performed with the statsmodels Python package \citep{seabold2010statsmodels}, using the lowess smoother with a data fraction of 0.5. All other parameters are set at their default values. The skew towards higher values of $\cos{(\delta)}$ \redtwo{at the simulated energies above 50 keV} indicates the model is able to extract useful information on the initial electron direction.}
    \label{fig:track_direction_reconstruction}
\end{figure}

The model is \redtwo{mostly} identical to that used in Section \ref{subsec:deterministic_model}. \redtwo{The difference is that the final 3 outputs shown Figure \ref{fig:3DCNNDiagram} are concatenated with the predicted electron track head position produced by the trained model from Section \ref{subsec:deterministic_model}, and then reduced with a feed-forward layer back to 3 numbers which predict the $X$, $Y$, and $Z$ components of the scattered electron's initial direction unit vector.} In principle the initial direction vector has only two independent variables. Removing the degeneracy of the third fitted parameter may further improve the direction reconstruction in future studies. \red{The model was implemented with Keras, using the Adam optimizer, a mean-squared-error loss function, and was trained for 20 epochs on an NVIDIA Tesla A100 GPU at the SLAC Scientific Data Facility, which took about 1 minute.} In Figure \ref{fig:track_direction_reconstruction} we show the performance of this model using a histogram of the cosine of the angle between the predicted electron direction and the true electron direction. A model that can reconstruct the direction with perfect accuracy would present as a delta function at 1 in this plot, while random guessing would produce a flat distribution. Since the distribution is skewed towards higher values of $\cos{(\delta)}$, this indicates it is able to extract useful information on the initial electron direction from the pixel data.

\begin{figure}
    \centering
    \includegraphics[width=\columnwidth]{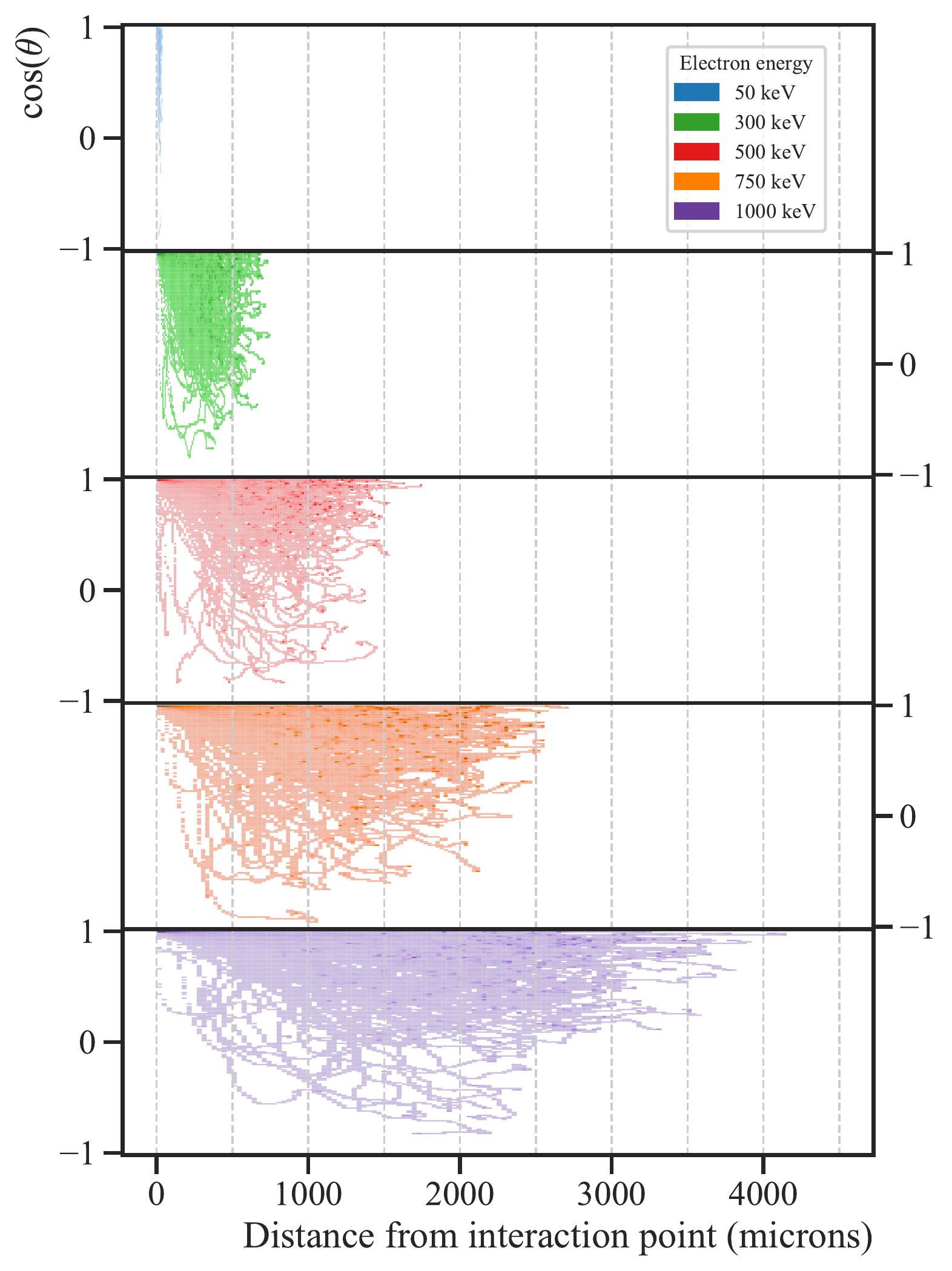}
    \caption{\redtwo{Each panel shows a 2D histogram of a collection of simulated electron tracks of a given energy, before any drifting or diffusion has been applied. For each track, we compute the absolute distance between the track origin and each step in the simulation for the $X$-coordinate, and for the $Y$-coordinate we compute the cosine between the electron's original direction and the vector between each pair of adjacent steps (i.e.~its direction of travel at each step). By definition, each track originates at $X=0$ and $Y=1$ in this diagram. In this way, we can show how far electrons tend to travel for each simulated energy, and how closely their directions of travel hew to their original direction as they travel. Tracks that curl back on themselves in real space will tend to travel downwards in these histograms, while straighter tracks will move mostly from left to right. The Bragg peaks at the ends of tracks are also visible in this plot; they appear as overdensities in the histogram, because the distance between steps decreases as $dE/dx$ increases in this regime. The vertical dashed lines show the simulated pixel pitch of 500 $\upmu$m. The color scale is linear.}}
    \label{fig:electron_scattering}
\end{figure}

It can be seen in Figure \ref{fig:track_direction_reconstruction} that the quality of the direction reconstruction increases with electron energy, as the $\cos{(\delta)}$ distributions peak more towards 1. Specifically, we find that the median angle between the true and predicted initial electron direction is \redtwo{87\degree, 71\degree, 59\degree, 50\degree, and 48\degree~for 50, 300, 500, 750, and 1,000 keV electrons respectively.}\footnote{\redtwo{A median angle of 90\degree\ would indicate no ability to reconstruct the direction, while a median angle of 0\degree\ would indicate perfect reconstruction.}} \red{This could be because high energy electrons travel in a straighter line initially}, making it easier to establish the initial direction the electron was traveling. \redtwo{This effect is visible in Figure \ref{fig:electron_scattering}, which shows the cosine of the angle between each electron's current and initial direction of travel, vs.\ the distance between its current and original positions. By comparing the 1,000 keV electrons (purple) and the 300 keV electrons (green), we see that the 1,000 keV electrons travel further before starting to deviate from their initial direction (i.e.~moving away from a $Y$-value of 1). This trend of higher-energy electrons traveling straight initially for longer than low-energy electrons holds up as a general trend in this energy regime.} This runs counter to the effect of energy on position reconstruction, where lower-energy scatters create more localized charge clouds, which facilitate better position reconstruction. The fact that these effects offset is encouraging, as it suggests that superior electron direction reconstruction for high-energy electrons could counteract lower-quality interaction position reconstruction.

\subsection{Uncertainty Quantification for Deep Learning Based Electron Track Head Reconstruction}
\label{subsec:edl_model}
Deep learning models are often considered to be high-fidelity models with accurate predictive powers. However, even predictions from a well-trained neural network may contain substantial errors and uncertainties due to bias, noise and complexity inherent in the data, the volume of the training data, the error minimum chosen during optimization, etc. In this light, it is important to be able to construct reliable prediction intervals in addition to point predictions. This can enable otherwise unavailable techniques when the data-driven model is to be used in scientific discovery. Reliable uncertainty quantification for deep learning has been identified as a priority for scientific machine learning (SciML) applications \citep{baker2019workshop, ihme2022combustion}. In our case, since electron tracks are highly random, they will have great diversity in shape, particularly at higher energies. It is reasonable to assume that the accuracy of a machine learning model may vary as a function of electron energy and/or track shape. \redtwo{The ability to select only tracks where the model has high confidence in its reconstruction should improve the overall pointing accuracy of the instrument.}

There exist different algorithms for uncertainty quantification for deep learning models. These include purely Bayesian approaches, non-Bayesian approaches and hybrid methodologies. \textit{A priori}, one is typically unable to determine the best algorithm for a specific application. As of yet, there have been very few studies focusing on evaluating, comparing and contrasting these approaches for scientific problems. This is exacerbated by the fact that most computer science studies comparing these algorithms have focused on classification problems, as opposed to problems requiring regression as is the case in many science applications, including this one. To this end, we evaluate different deep learning uncertainty quantification approaches on the data sets for electron track head reconstruction. These include Bootstrapped Ensembles of neural networks, Monte Carlo Dropout, Probabilistic Neural Networks (PNNs), and Evidential Deep Learning, a brief description of which follows. \red{Additionally, we also outline the hyperparameters of the algorithm and their selection using the validation dataset.} 

Bootstrapped ensembles use bootstrap aggregation for uncertainty estimation. Bootstrap data sets of $N$ samples each are produced for training each individual model in the ensemble by sampling uniformly with replacement from the original training data set of $N$ unique samples. Individual neural networks in the model ensemble are trained separately on each of these bootstrap data sets. The predictions of the individual trained models in the ensemble are averaged to get a mean prediction. The variance in predictions provides a measure of the predictive uncertainty. Boostrapped ensembles are very commonly used in uncertainty quantification for scientific problems \citep{manly2018randomization, jeong199321}. \red{Herein, the key hyperparameter is the number of constituent neural network models in the ensemble. In this investigation, the number of neural network models in the ensemble was increased until the coverage and MPIW became constant over the validation dataset, beyond a value of $15$ individual neural network models in the ensemble. Thus, the comparison and the comparative results correspond to a bootstrapped ensemble with $15$ neural network models.}

Monte Carlo Dropout is a Bayesian technique introduced in \cite{gal2016}, where one approximates the network's posterior distribution of class predictions by collecting samples obtained from multiple forward passes of dropout regularized networks. \textit{Dropout regularization} \citep{dropout} involves random omissions of feature vector dimensions during training time, which is equivalent to masking rows of weight matrices. Inclusion of dropout layers mitigates model overfitting and is empirically known to improve model accuracy \citep{dropout}. A key observation of \cite{gal2016} is that under suitable assumptions on the Bayesian neural network prior and training procedure, sampling $N$ predictions from the BNN's posterior is equivalent to performing $N$ stochastic forward passes with dropout layers fully activated. In this manner, the full posterior distribution may be approximated by Monte Carlo integration of the posterior predicted probability vector. Monte Carlo Dropout networks may also be interpreted as a form of ensemble learning \citep{dropout}, where each stochastic forward pass corresponds to a different realization of a trained neural network. \red{In the usage of MC Dropout, key hyperparameters include the placement and number of dropout layers, the strength of the dropout rate \citep{kendall2015bayesian, devries2018leveraging}, the number of forward passes sampled during prediction, etc. During model selection, the best performance over the validation set was engendered by the architecture utilizing a dropout layer after each convolutional block and a single dropout layer after the global average pooling operation. Explicitly, in this architecture, no dropout was performed between the input and the first convolutional layer and neither between the last dense layer and the output. Additionally, in between the convolutional layers the best dropout rate was $0.25$ and in the dense layers the best dropout rate was $0.1$. All the hyperparameters were selected using cross-validation. While using higher dropout rates led to larger prediction intervals, they also had a deleterious effect on the model accuracy leading to poorer calibration of the uncertainty estimates. Beyond 50 samples for prediction, we did not observe any improvement in the metrics and thus, 50 samples in the prediction were used.} 

Probabilistic Neural Networks (PNNs) use a proper scoring rule \citep{gneiting2007strictly} to evaluate their performance. For regression tasks, this scoring rule is the negative log likelihood (NLL). Consequently, the training of the model ascertains network parameters that give an optimal mapping from the space of raw features to the mean and standard deviations of the predictions, so as to minimize the NLL over the training data set. \redtwo{Strictly speaking, this approach entails no additional hyperparameters} and only requires a change to the cardinality of the output and the usage of the NLL as the loss function to be optimized. However, the learning rate of the Adam optimizer was adjusted to ensure optimal convergence for this algorithm as well.

Evidential Deep Learning (EDL) \citep{amini2020deep}, refers to a class of deep neural networks that exploit conjugate-prior relationships to model the posterior distribution analytically. EDL methods have the immediate advantage of requiring only one single pass to access the full posterior distribution, at the price of restricting the space of possible posterior functions to that containing only those which have the appropriate conjugate prior forms. They also only require one to modify the loss function and the final layer of its deterministic baseline, which allows flexible integration with complex, hierarchical deep neural architectures. \redtwo{Under the assumption that the input data are drawn from a Gaussian distribution of unknown mean and standard deviation, the conjugate prior is then a normal-inverse-gamma distribution. In our case, we place an independent prior on each Cartesian coordinate of the electron track head position.} \red{For EDL-based regression, the hyperparameter to be determined is the coefficient for the regularization loss defined by Equation 9 of \cite{amini2020deep} (thus, for EDL, total loss is the sum of the NLL loss and the product of the regularization coefficient with the regularization loss). While comparing the EDL model to the other UQ models, the regularization coefficient was utilized at a value of 1.}

\begin{figure*}
    \gridline{
        \hfill
        \vbox{
            \parskip=0pt
            \hsize=\columnwidth
            \includegraphics[width=\columnwidth,     trim={5.3cm 2.2cm 3.6cm 1.45cm},clip]
                {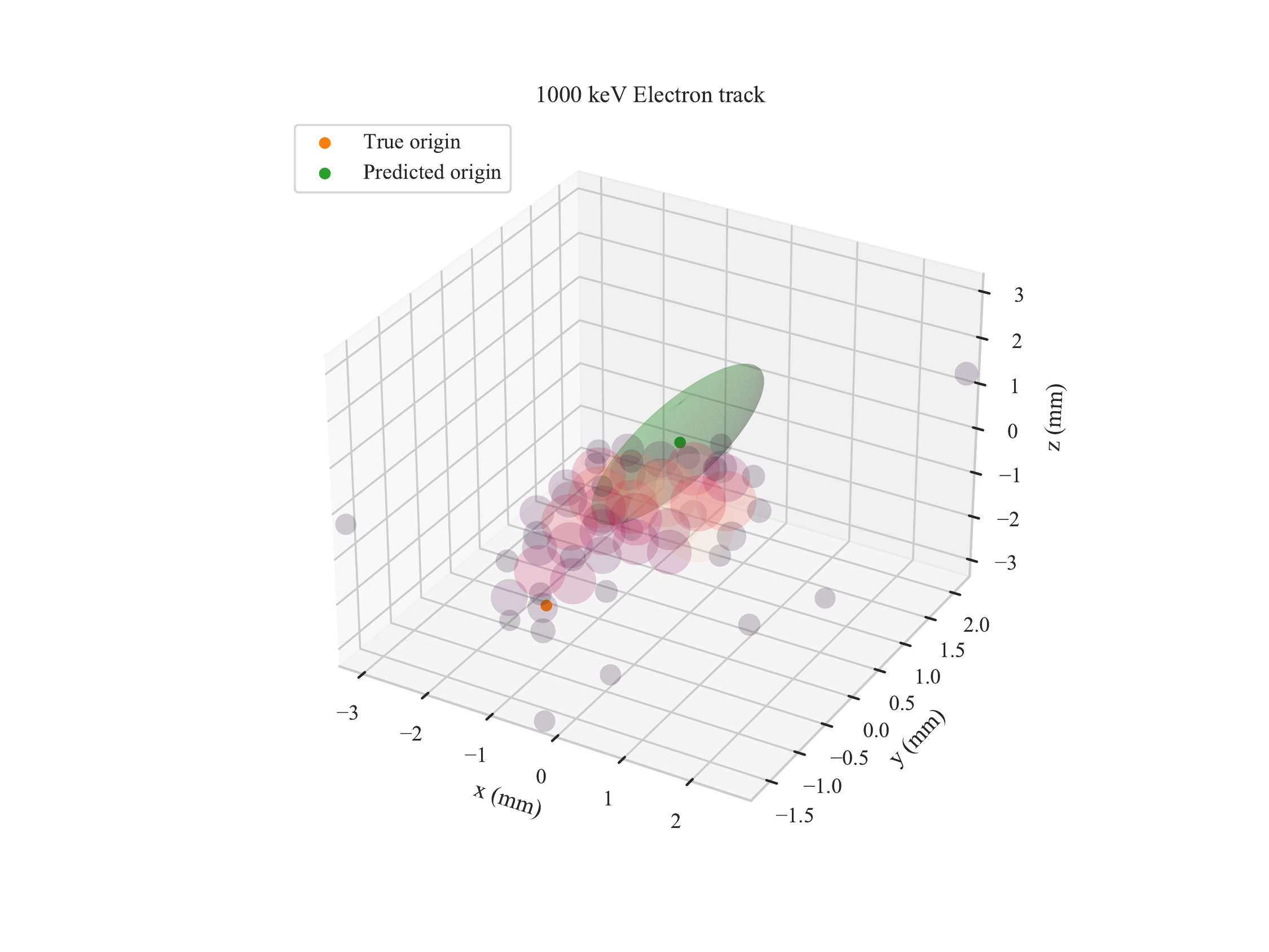}
            \vskip2pt\vtop{
                \footnotesize
                \hsize=\columnwidth
                (a) Highest error track at 5 cm drift.\vskip1pt
            }
        }
        \hfill
        \hfill
        \vbox{
            \parskip=0pt
            \hsize=\columnwidth
            \includegraphics[width=\columnwidth,     trim={5.3cm 2cm 5.1cm 1.45cm},clip]
                {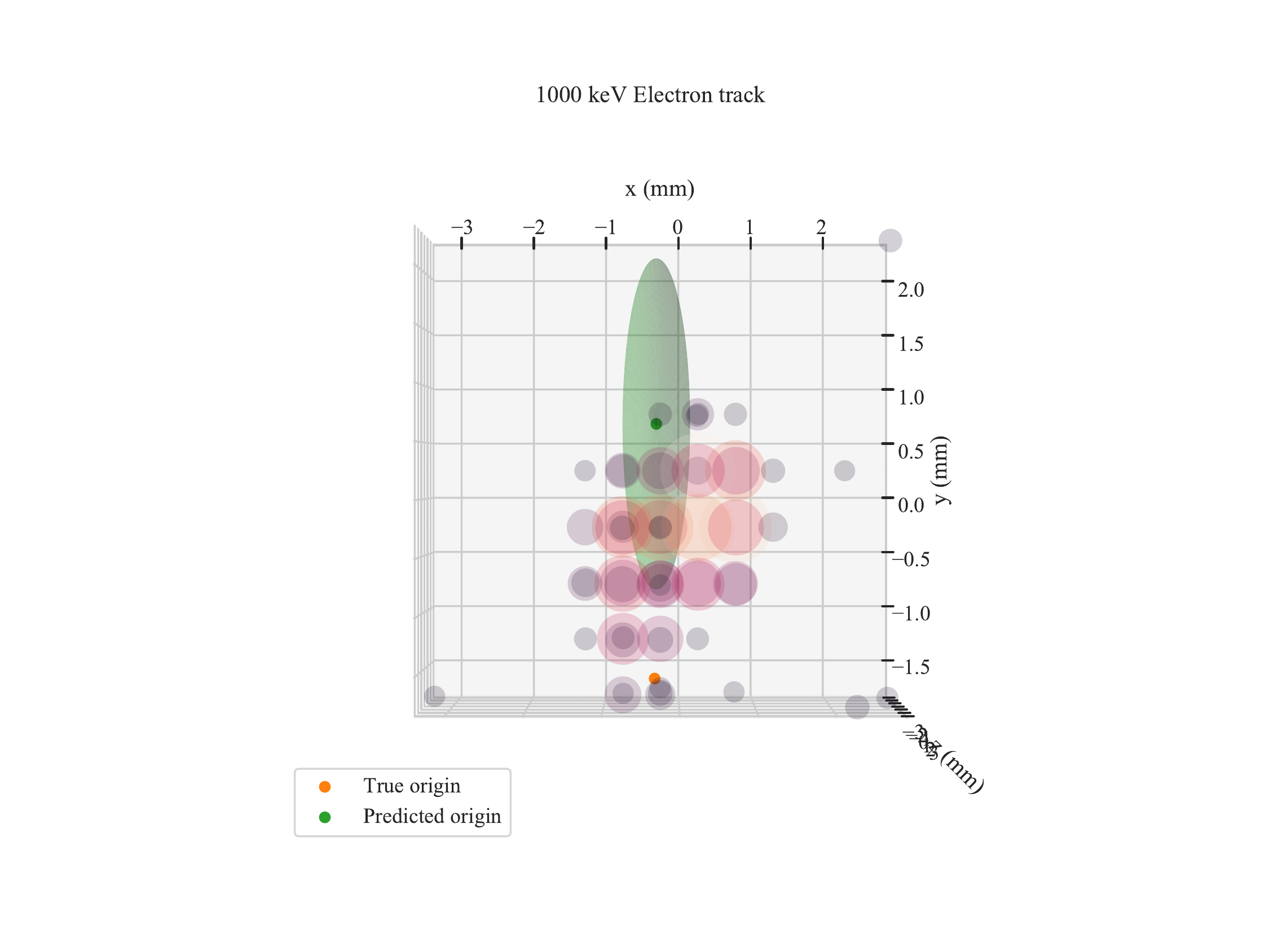}
            \vskip2pt\vtop{
                \footnotesize
                \hsize=\columnwidth
                (b) A view along the $Z$-axis of the highest error track at 5 cm drift.\vskip1pt
            }
        }
        \hfill
    }
    \gridline{
        \hfill
        \vbox{
            \parskip=0pt
            \hsize=\columnwidth
            \includegraphics[width=\columnwidth,     trim={7.4cm 4.1cm 4.5cm 1.45cm},clip]
                {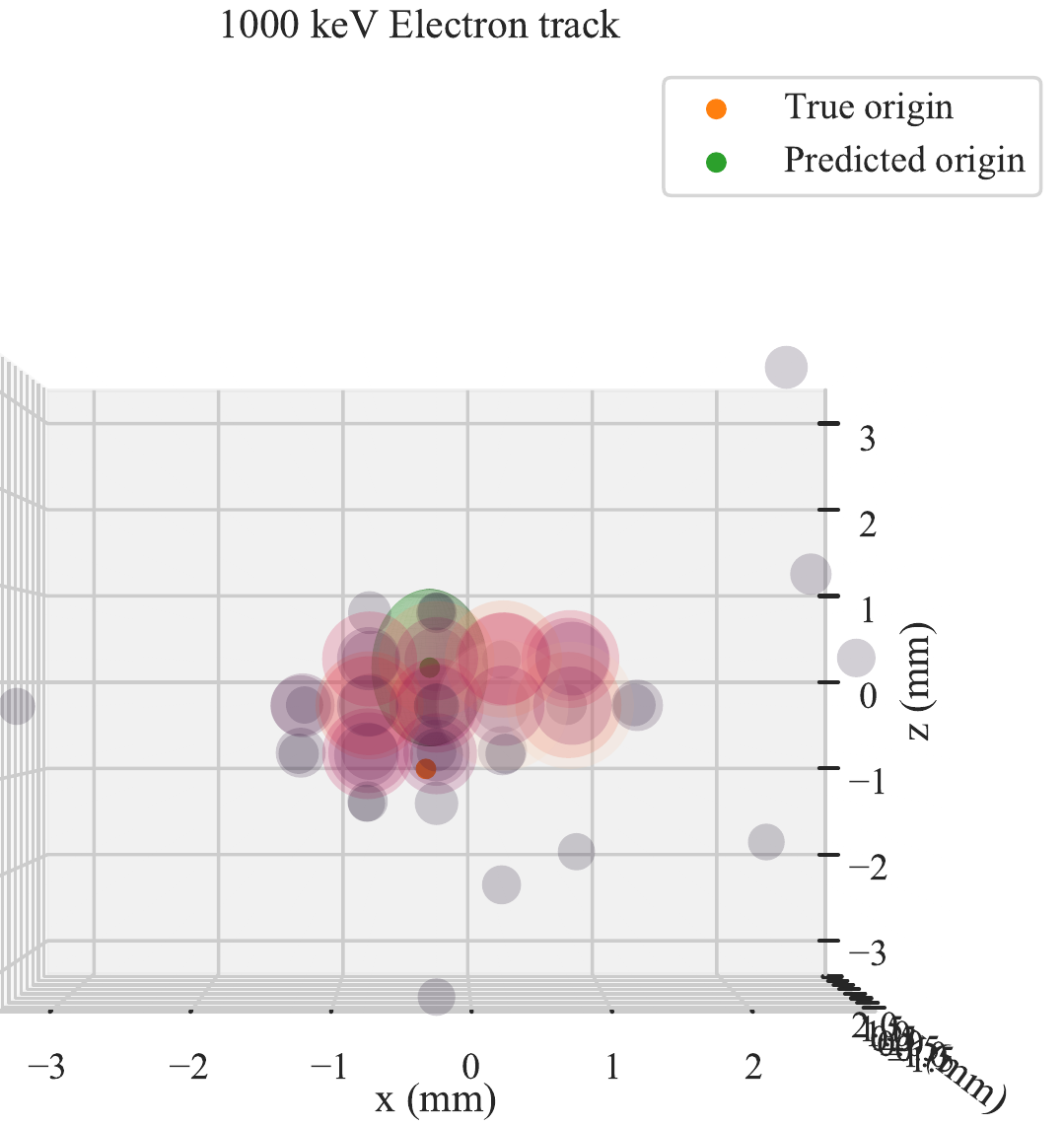}
            \vskip2pt\vtop{
                \footnotesize
                \hsize=\columnwidth
                (c) A view along the $Y$-axis of the highest error track at 5 cm drift.\vskip1pt
            }
        }
        \hfill
        \hfill
        \vbox{
            \parskip=0pt
            \hsize=\columnwidth
            \includegraphics[width=\columnwidth,     trim={5.3cm 2cm 3.6cm 1.45cm},clip]
                {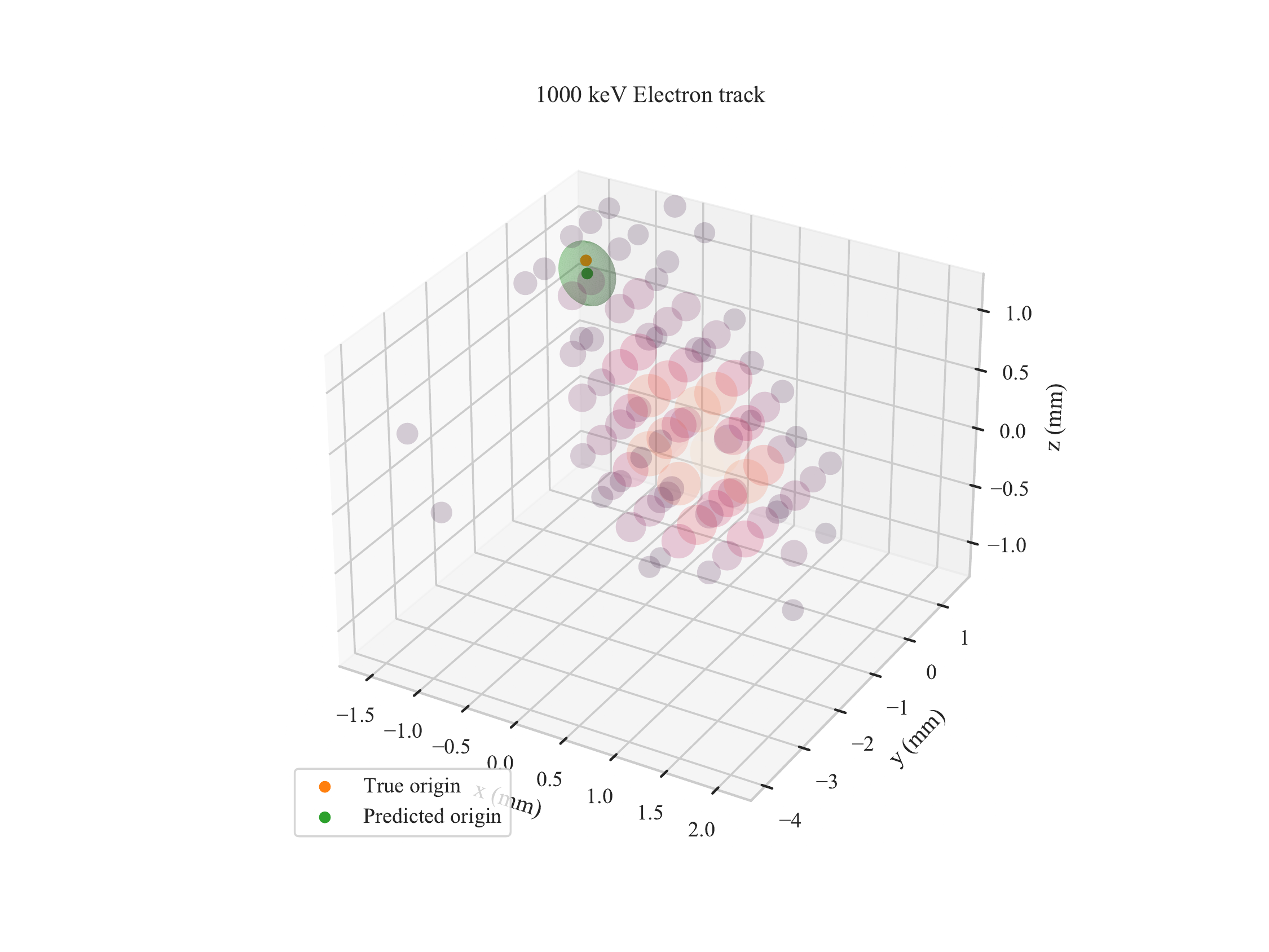}
            \vskip2pt\vtop{
                \footnotesize
                \hsize=\columnwidth
                (d) The lowest error track at 20 cm drift.\vskip1pt
            }
        }
        \hfill
    }
    \caption{Two chosen examples of simulated detector readouts for 1,000 keV electron tracks with true and reconstructed head locations included, as well as a 1$\sigma$ elliptical uncertainty boundary on the reconstructed position, given by the green ellipsoid. Each pixel above threshold is represented by a circle, with its size and color proportional to the energy collected.}
    \label{fig:track_examples}
\end{figure*}

To ascertain the quality of the uncertainty estimates from the different algorithms, we compute three metrics: accuracy, coverage and sharpness. Accuracy denotes the fidelity of the mean prediction, evaluated via the RMS error on the test data set. Coverage \citep{romano2019conformalized} measures the calibration of the uncertainty estimates. The prediction interval is the estimate provided by a model of the range in which the target value will lie with a certain probability, given the input features. If the actual value lies in the predicted interval, it is said to be covered by the interval. Coverage measures the ratio of test samples that lie in the prediction interval of a specific quantile. A critical consideration for prediction intervals is their reliability at reflecting the quantiles in the distribution of the target quantity. \redtwo{As an illustration, a $95\%$ prediction interval that subsumes less than $95\%$ of the target samples will lead to predictions with higher error than expected, which could have significant consequences, depending on the application.} Prediction intervals with incorrect coverage indicate poorly calibrated models, where calibration here indicates the degree to which the predicted uncertainty matches the true underlying uncertainty in the data. \red{In our comparison between algorithms in Table 1, we utilize a uniform prediction interval of $95\%$ for all the algorithms and report the coverage at this prediction interval.} Finally, sharpness measures the size of the prediction interval, at a fixed value of coverage. We measure the sharpness of the prediction intervals using the Mean Prediction Interval Width (MPIW) \citep{pearce2018high}, which is the average size of the prediction intervals over the test data set. We require that the prediction intervals provide a minimum value of the coverage, while maintaining the MPIW to low values. To ensure a fair comparison, we compute the MPIW at a fixed coverage, by increasing the prediction interval until a set coverage is reached before making the calculation. Specifically, we measure MPIW values at prediction intervals that ensure $95\%$ coverage over the test data set. 

The evaluation of deep learning uncertainty quantification algorithms for the electron head reconstruction is reported in Table~\ref{tab:table1}. We observe that the ensemble-based approaches lead to a decrease in prediction error over the single deterministic model. This is due to the aggregation over decorrelated individual models and forms the basis of bagging-based meta-models \citep{zhou2021ensemble}. However, the variance between the predictions of individual models in the ensemble is not a reliable indicator of predictive uncertainty. As an illustration, we find that the Bootstrapped ensembles are leading to a highly overconfident estimate of the predictive uncertainty with a coverage of only $0.22$ for a $95\%$ prediction interval. To this end, we would caution against reliance on uncertainties from Bootstrapped ensembles in use cases similar to this one. The Evidential Deep Learning approach gives well calibrated uncertainties, where the coverage is $0.79$ for a $95\%$ prediction interval. The mean prediction interval width generated by the Evidential Deep Learning models is moderated as well. This indicates that the EDL models provide sharpness of the predicted intervals. Thus, based on this empirical study, we utilize the Evidential Deep Learning approach for uncertainty estimation in this investigation. \red{After selection of the EDL approach, we tuned the regularization coefficient to 0.02 to enforce correct coverage. The model uses the Evidential Deep Learning for Regression framework \citep{amini2020deep}. Herein the model architecture is identical as the prior model in Section \ref{subsec:deterministic_model}. However, the final layer outputs the parameters of the posterior distribution on the position of the head, as opposed to the actual estimate for the position of the head. The final model was trained using the Adam optimizer with a learning rate of $5\cdot 10^{-5}$ over 1500 epochs. It was trained at the SLAC Scientific Data Facility on an NVIDIA Tesla A100 GPU, which took about 90 minutes.}

\begin{table}

\centering

\caption{\label{tab:table1} Evaluation of Different Deep Learning Uncertainty Quantification Algorithms}

\begin{tabular}{cccc}
\tableline \tableline
Algorithm& RMSE & Coverage& MPIW\\\tableline
Single Deterministic Model& 0.43& --& --\\
Bootstrapped Ensembles& 0.39& 0.22& 4.75\\
MC Dropout& 0.40& 0.46& 4.33\\
PNN& 0.42& 0.41& 4.47\\
Evidential Deep Learning& 0.40& 0.79&3.35\\\tableline \tableline
\end{tabular}

\end{table}

Figure \ref{fig:track_examples} shows several examples of simulated 1,000 keV electron tracks. The amount of charge in each pixel is given by the colored scatter plot of circles, where the size and color of each circle is proportional to the charge signal. The true and reconstructed origin of the electron track are given by the orange and green dots respectively, with an ellipsoid around the reconstructed origin representing the predicted uncertainty. The examples have been chosen to illustrate different possible outcomes of the reconstruction. \redtwo{Panels \ref{fig:track_examples}a, \ref{fig:track_examples}b, and \ref{fig:track_examples}c show three different views of the event with the largest absolute reconstruction error. Here the model appears to have identified the wrong end of the track as the head. However, it has compensated for this mistake somewhat by correctly estimating a large error on the $Y$ dimension. While perhaps not obvious from these examples, it is the case that the uncertainty ellipsoids have no diagonal components. This is because the 3 cardinal directions are treated independently by the model, and so it lacks the ability to include non-diagonal components in the covariance matrix for the uncertainty, leading to ellipsoids with axes always aligned with the cardinal axes. Figure \ref{fig:track_examples}d shows the 20 cm drift track with the smallest reconstruction error. Here we clearly see the Bragg peak in the lower center of the image (larger/lighter circles), and the model has correctly identified that the opposite end of the track is the true track head.}

\begin{figure}
        \includegraphics[width=\columnwidth, trim={0.9cm 0.7cm 0.75cm 0.65cm},clip]{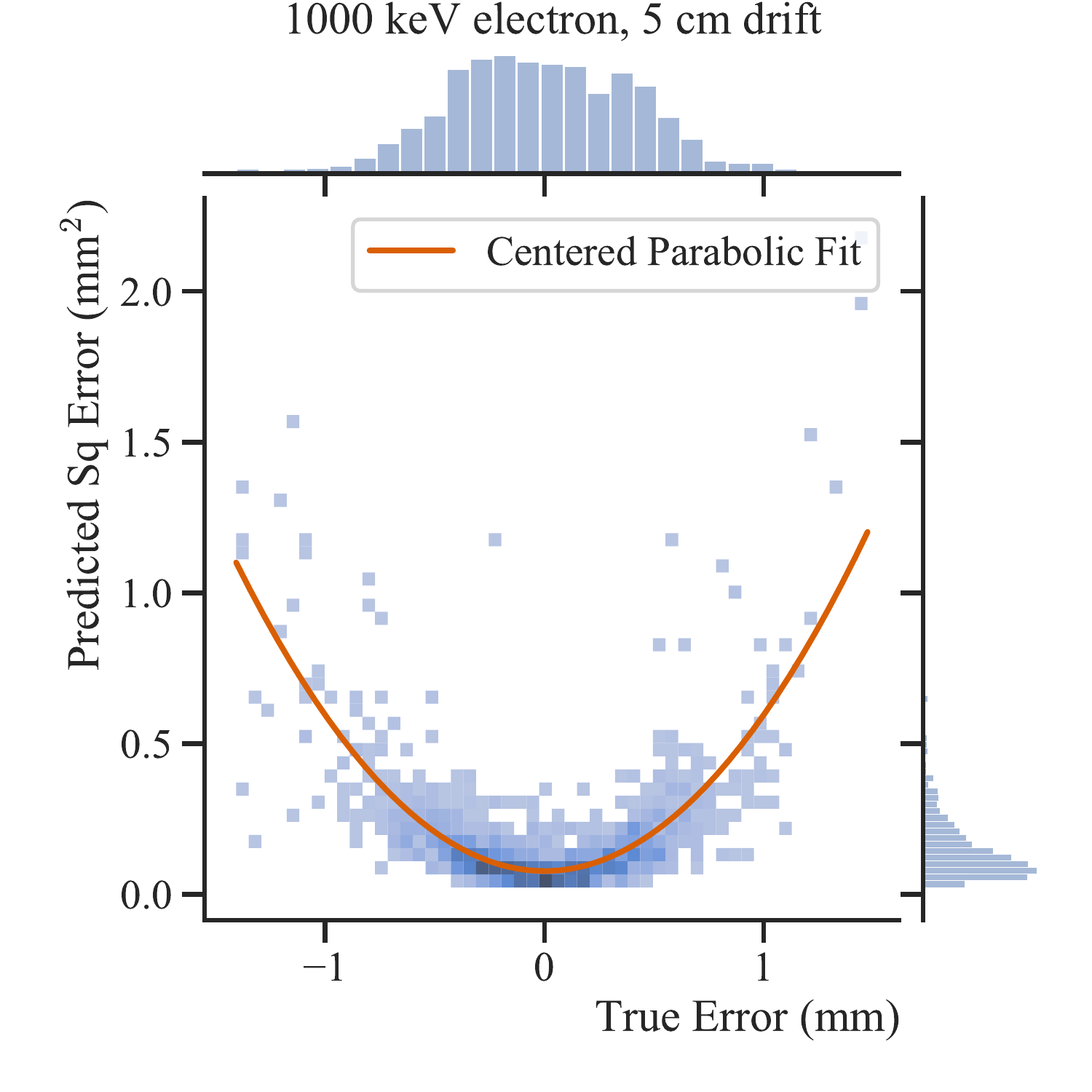}
        \includegraphics[width=\columnwidth, trim={1.0cm 0.7cm 0.75cm 0.65cm},clip]{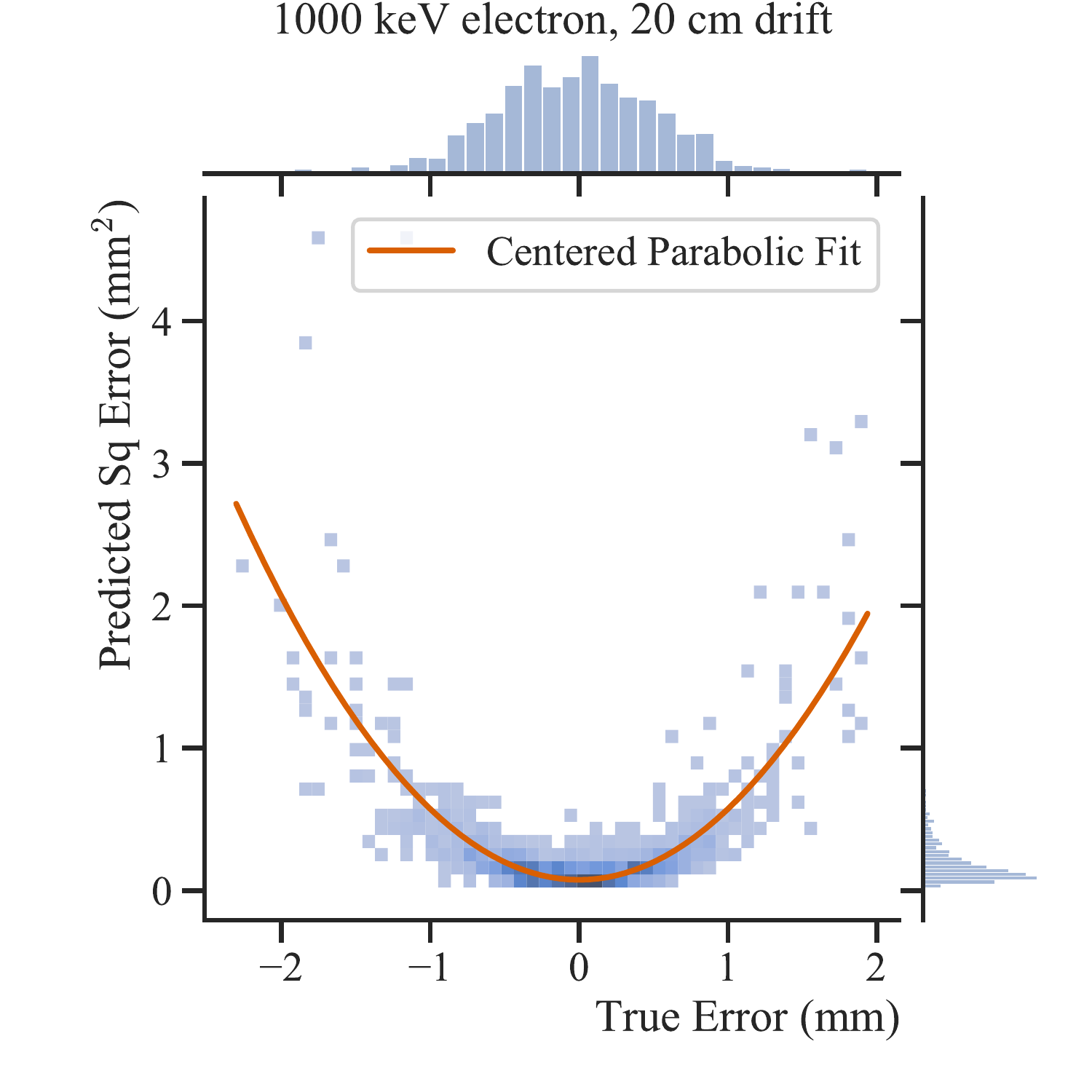}
    \caption{Predicted squared error vs.\ true error for 1,000 keV electron tracks at 5 cm drift (top) and 20 cm drift (bottom). A model that perfectly estimates its uncertainty would lie on a parabola.}
    \label{fig:unc_vs_error}
\end{figure}

\redtwo{To verify the model's uncertainty calibration, it is instructive to plot the output of the model in a 2D space: the model-predicted squared error on the electron track head position against the true prediction error (i.e.~the difference between the true and predicted position of the electron track head). As can be seen in Figure \ref{fig:unc_vs_error}, these two parameters have a parabolic relationship, and achieve a small non-zero minimum value for the predicted squared error at a true error of zero. This is expected, since the posterior distribution of the predicted squared error is inverse-gamma, which approaches zero exponentially as its variable approaches zero. This can be seen in the marginal Y-axes of Figure \ref{fig:unc_vs_error}. As can also be seen in the marginal X-axes of Figure \ref{fig:unc_vs_error}, the 1D distribution of the true error is normally distributed, as expected given the normal-inverse-gamma posterior distribution of the EDL loss function.}

\begin{figure}
    \centering
    \includegraphics[width=\columnwidth, , trim={0.6cm 0.7cm 0.6cm 0.65cm},clip]{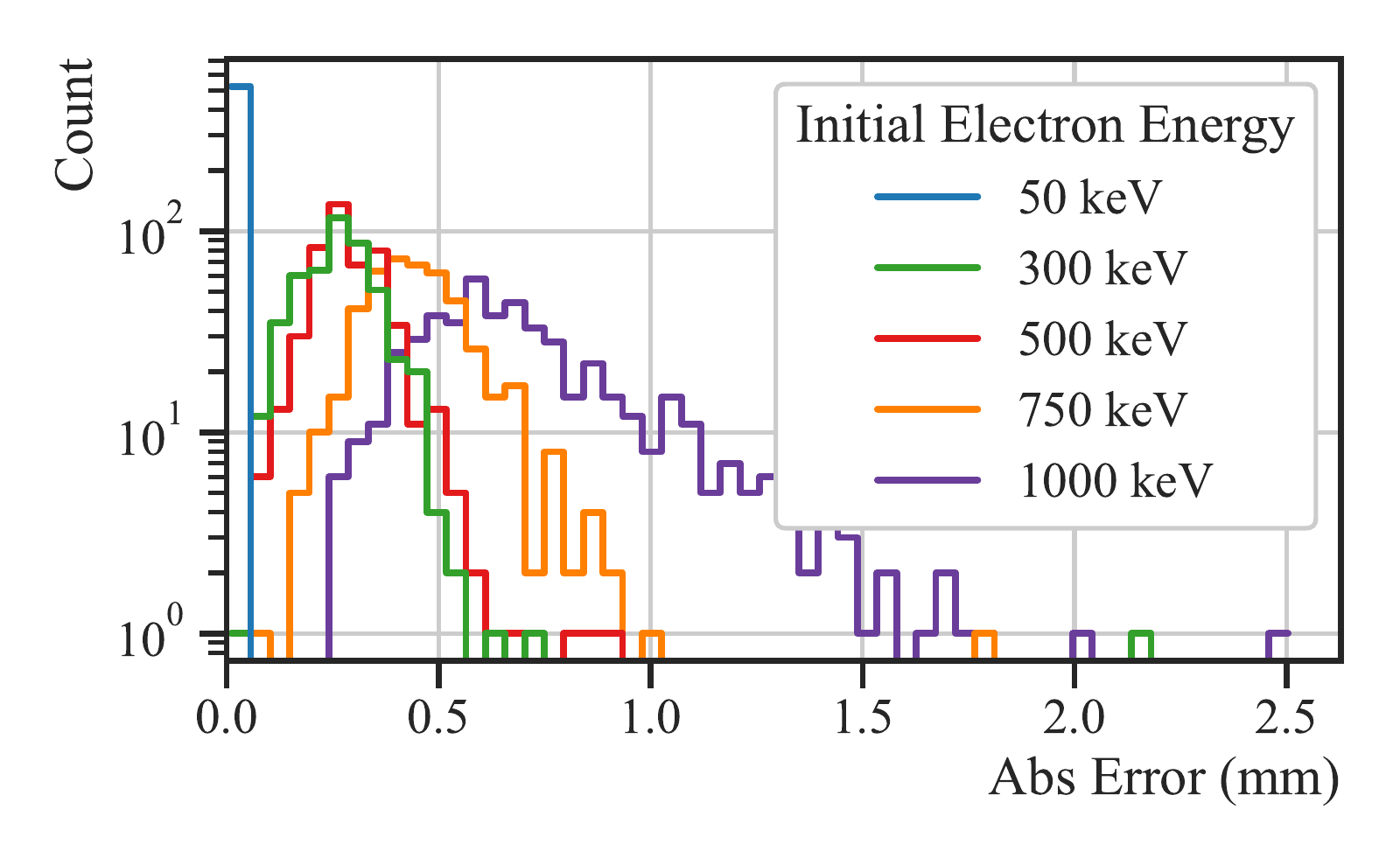}
    \caption{Empirical absolute error distributions of the trained EDL model, broken down by initial electron energy. \redtwo{This figure gives the actual distance} between the predicted and true electron track head locations.}
    \label{fig:edl_5_cm_drift_abs_error}
\end{figure}

\begin{figure}
    \centering
    \includegraphics[width=\columnwidth, trim={0.7cm 0.7cm 0.6cm 0.65cm},clip]{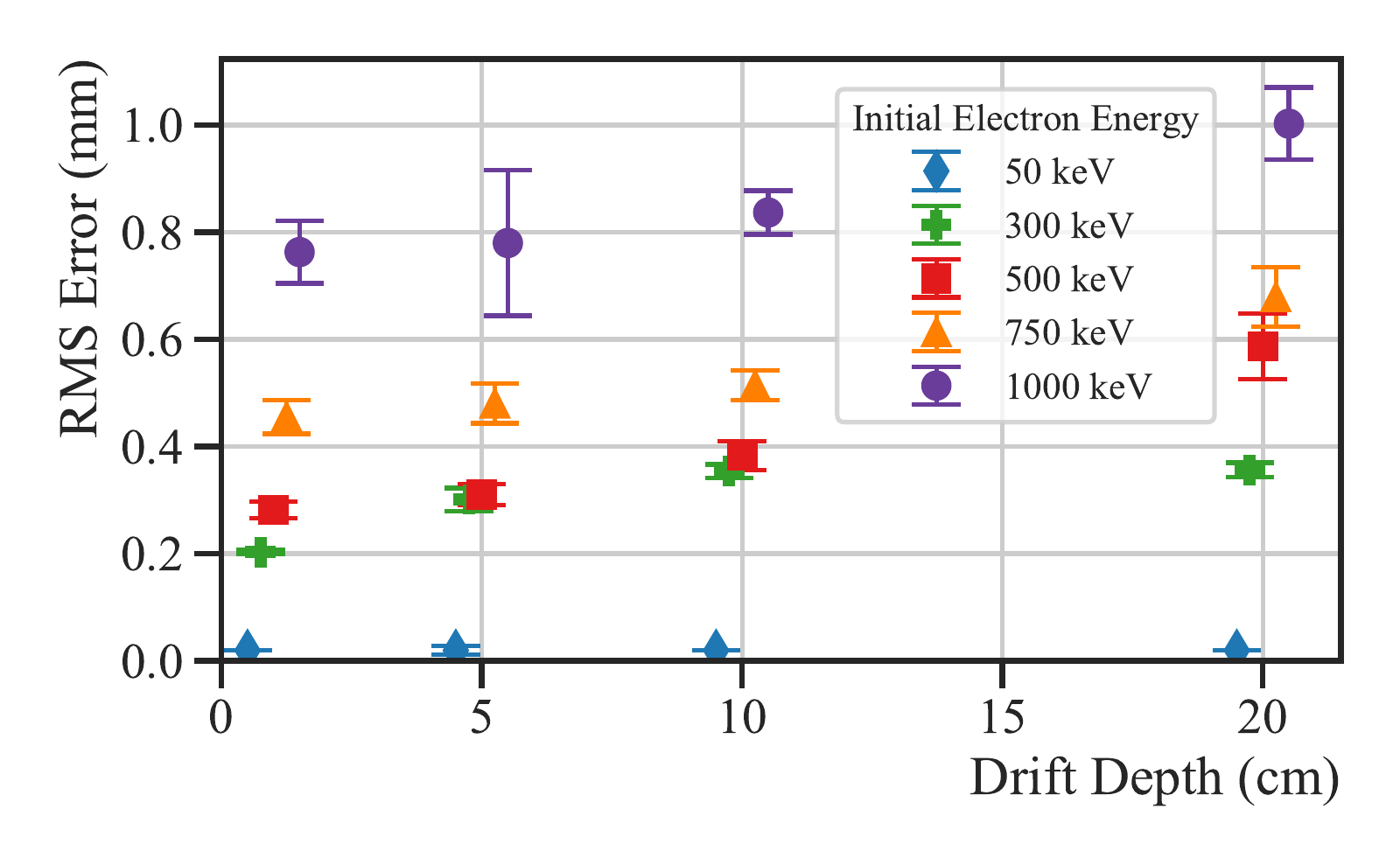}
    \caption{The \redtwo{square root of the mean of the distribution of the squared differences} between the predicted and true track head location for different initial electron energies as a function of drift depth, for the EDL model. We have added small $X$ offsets to the data points purely for visualization purposes. All of the points truly lie at 1, 5, 10, or 20 cm drift. The error bars are based on the model-estimated uncertainties, an example of which is the y-axis of Figure \ref{fig:unc_vs_error}. The uncertainty for each point is set by the width of the samples around the parabola in the relevant version of Figure \ref{fig:unc_vs_error}. If all of the samples in Figure \ref{fig:unc_vs_error} laid exactly upon the orange parabola, then the error shown in this Figure for that data point would be 0. This would indicate that the model was able to determine for each track exactly what the error in the track location was. This is however not the case, so the error bars in this figure indicate to what degree the RMS errors are due to unmodeled variation in track shape. A more detailed explanation of how this uncertainty is calculated is presented in the text.}
    \label{fig:edl_fwhm_vs_drift}
\end{figure}

Similar to Figure \ref{fig:deterministic_5_cm_drift_abs_error}, Figure \ref{fig:edl_5_cm_drift_abs_error} shows the distribution of absolute errors for the trained EDL model at a 5 cm drift depth, broken down by initial electron energy. We show the dependence on drift depth of the RMS error for each initial electron energy in Figure \ref{fig:edl_fwhm_vs_drift}. Since the EDL approach provides us with a modeled uncertainty, we can turn that into an estimate of the systematic error on the points in Figure \ref{fig:edl_fwhm_vs_drift}. The error bars on the points in Figure \ref{fig:edl_fwhm_vs_drift} are an estimate of the variation in electron track shapes, leading to incorrect location predictions, that is not modeled by the EDL approach. Unmodeled track shape variation can arise not only from imperfect design/training of the model, but also from variation at a scale too small to be distinguishable by the model, once the simulated drift and detector response has been applied. While there does appear to be a clear increasing trend of the RMSE as the drift length increases, the fluctuations about that trend mostly seem to be within the estimated systematic uncertainties. For each point, its error bars are constructed by 1) fitting inverse-gamma distributions to slices of the version of Figure \ref{fig:unc_vs_error} with the appropriate initial electron energy and drift depth, 2) computing the standard deviation of those inverse-gamma distributions, 3) computing the mean standard deviation across the slices, weighted by the number of samples in each slice, 4) using that mean standard deviation to compute the uncertainty on the mean squared error by dividing it by $\sqrt{n}$ where $n$ is the number of samples, and finally 5) propagating the uncertainty on the mean squared error through to the root mean squared error. The statistical uncertainty on the points in Figure \ref{fig:edl_fwhm_vs_drift} is negligible. The RMS error increases as a function of drift depth for all electron energies.

\begin{figure}
    \centering
    \includegraphics[width=\columnwidth, trim={0.68cm 0.7cm 0.7cm 0.65cm},clip]{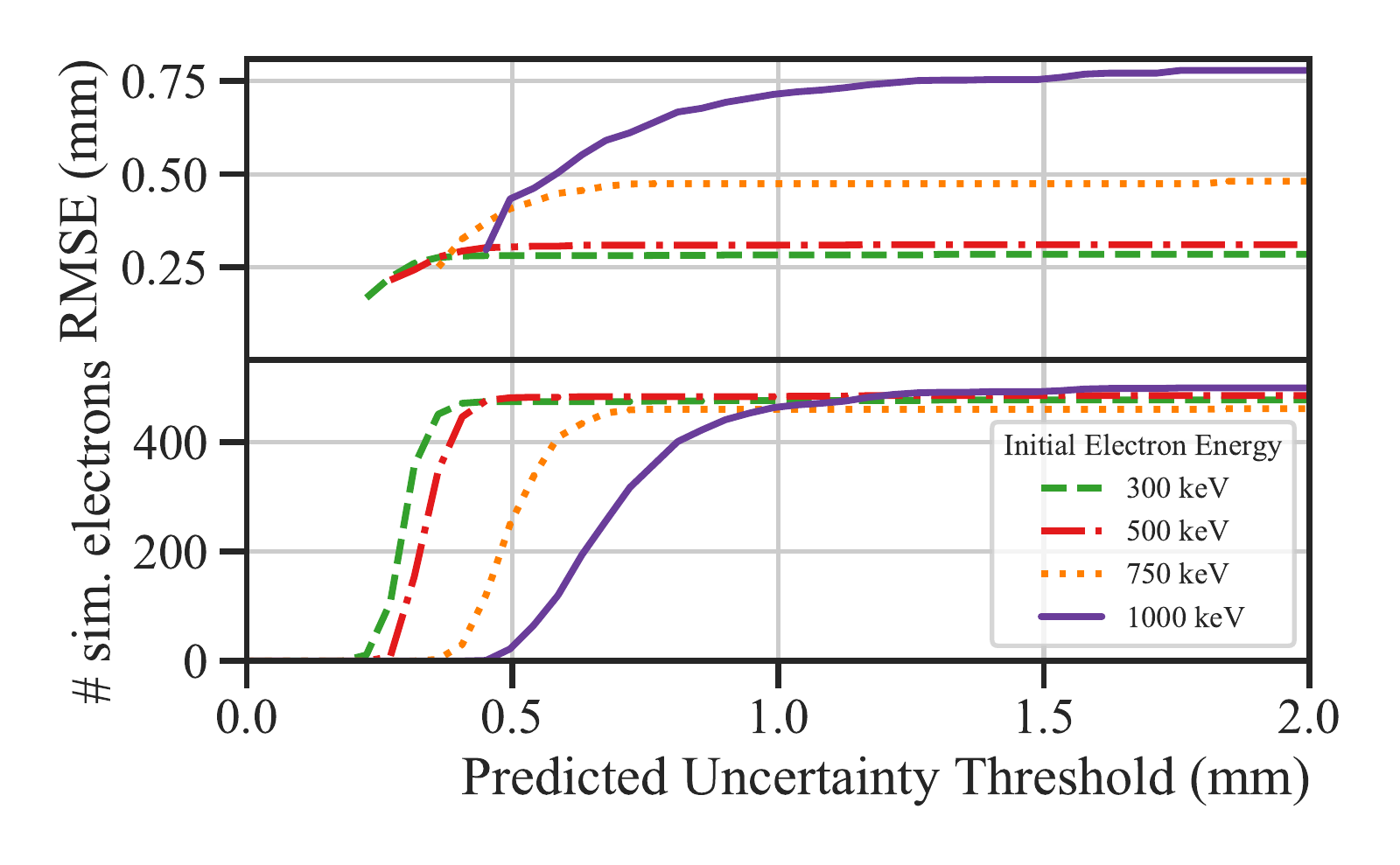}
    \caption{Root mean squared error on the reconstructed position as a function of uncertainty threshold for various initial electron energies at a 5 cm drift length.}
    \label{fig:unc_threshold_effects}
\end{figure}

Figure \ref{fig:unc_threshold_effects} shows the RMS error as function of an applied threshold in the predicted uncertainty, meaning that electron tracks with an uncertainty on the reconstructed track head position greater than the threshold are excluded from the calculation of the RMS error. The top panel shows the effect of excluding these tracks on the RMS error, while the bottom panel shows how many events are \redtwo{retained} as a function of the uncertainty threshold. For reference (as seen in Figure \ref{fig:det_fwhm_vs_drift}), \red{in the deterministic model} the RMS error for 1,000 keV electron tracks at a 5 cm drift depth is \redtwo{slightly more than 0.6 mm}. According to Figure \ref{fig:unc_threshold_effects}, \red{an RMS error similar to that of the deterministic model} can be obtained by placing a threshold of approximately 0.6 mm on the predicted uncertainty. However, this requires rejecting approximately 75\% of all 1,000 keV electron tracks.

\begin{figure}
    \gridline{\fig{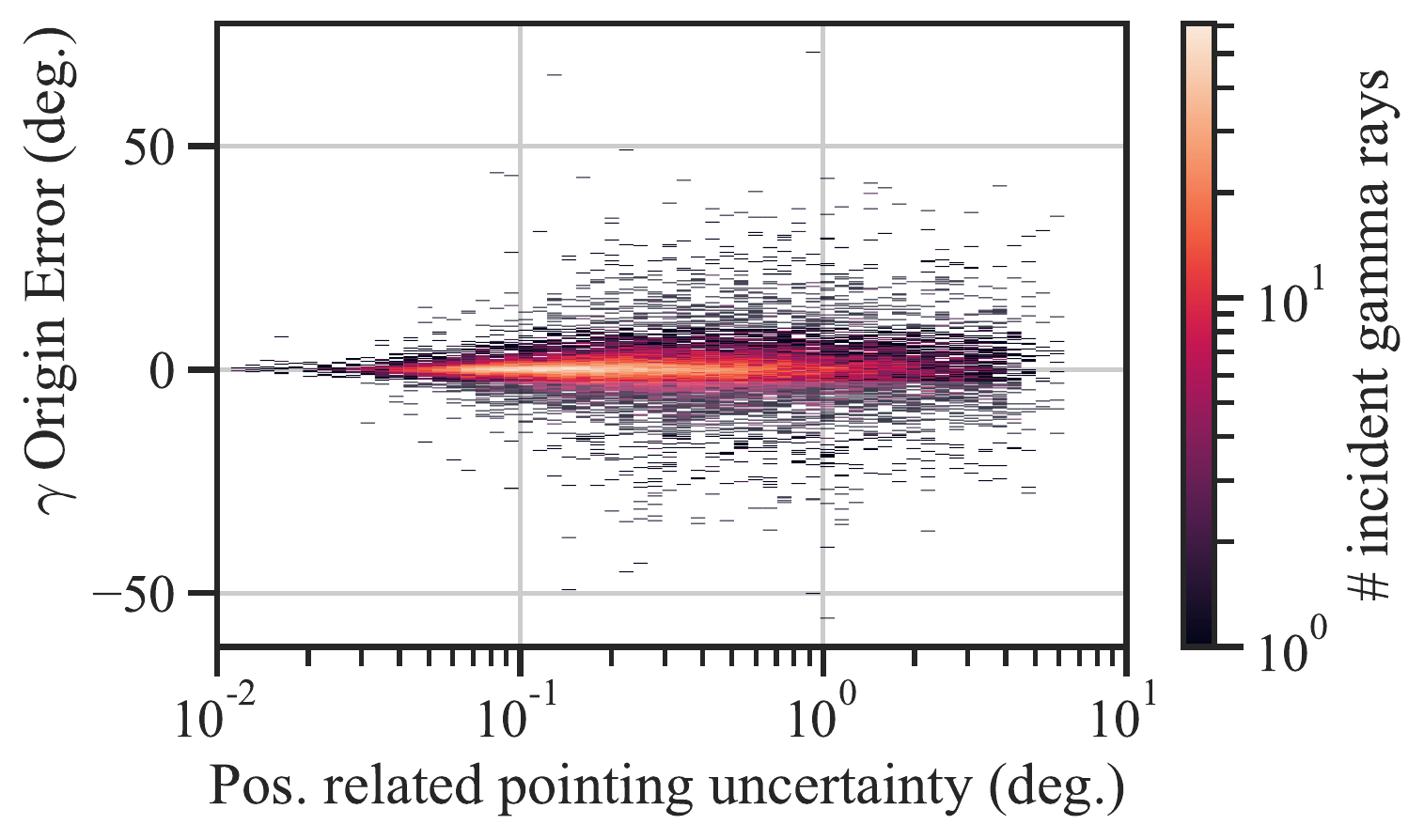}{\columnwidth}{(a) True gamma origin error vs.\ \redtwo{the component of the predicted uncertainty in the gamma-ray original direction due to uncertainty in the interaction location}}}
    \gridline{\fig{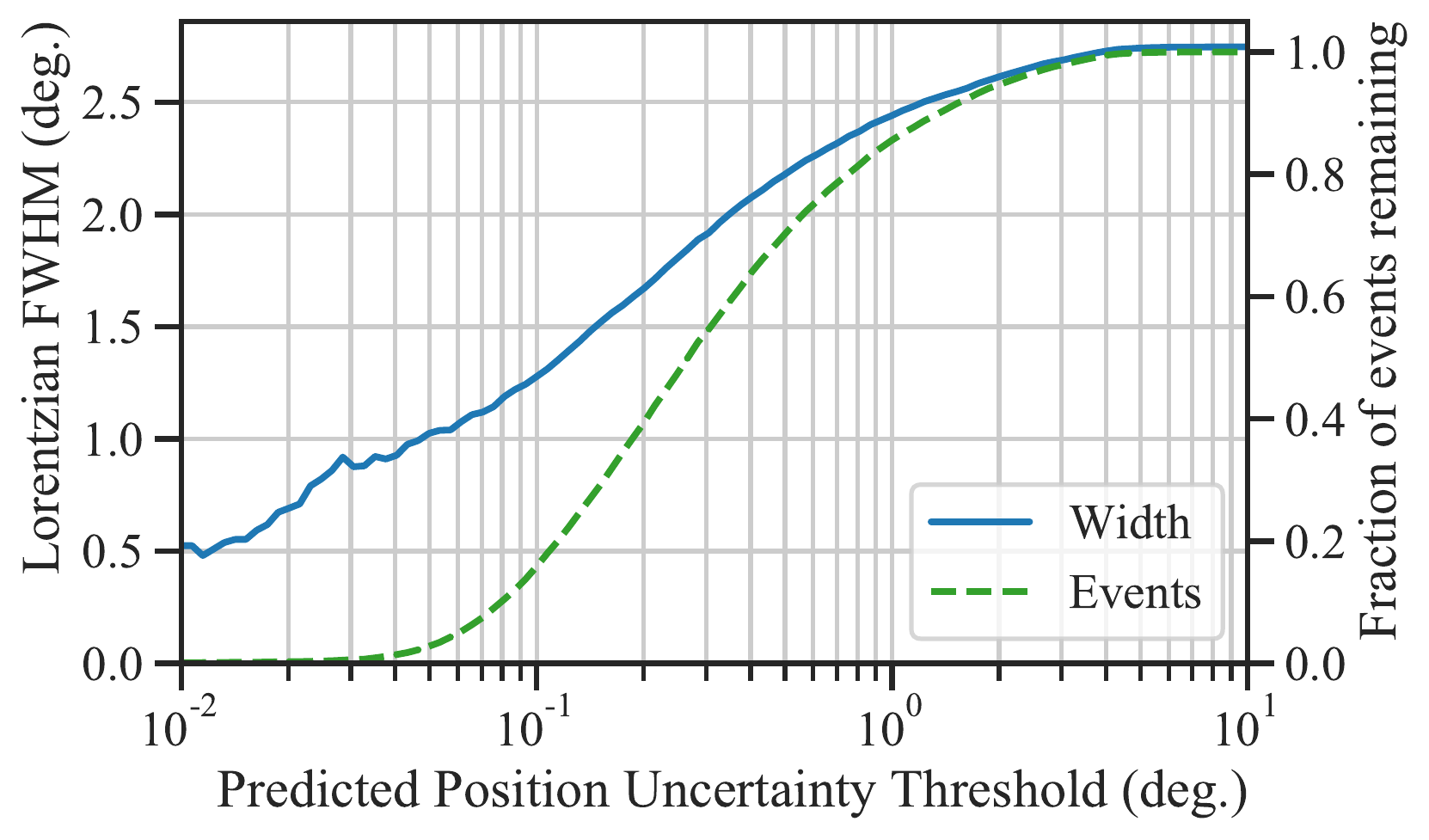}{\columnwidth}{(b) \redtwo{The projection of Figure \ref{fig:pos_unc_effects}a onto its $Y$-axis will produce an ARM plot like that in Figure \ref{fig:det_ARM}. This figure shows the FWHM of a fit of a Lorentzian to that ARM, as an upper-limit threshold is placed on the position related uncertainty in Figure \ref{fig:pos_unc_effects}a.} The solid blue curve gives the FWHM of the fitted Lorentzian as a function of the threshold, while the dashed green curve shows the fraction of events remaining for a given threshold.}}
    \caption{Using only tracks where the ordering of the individual Compton scatters was correctly reconstructed, we see that placing a threshold on the uncertainty in the reconstructed gamma-ray origin ring opening angle can reduce the width of the error distribution.}
    \label{fig:pos_unc_effects}
\end{figure}

Having verified the model's calibration, we know that removing events with high predicted uncertainty from the analysis should improve the pointing accuracy of the instrument. This effect can be seen in Figure \ref{fig:pos_unc_effects}. Figure \ref{fig:pos_unc_effects}a shows the true error of the reconstructed gamma-ray initial direction against the \redtwo{component of the pointing uncertainty that is attributable to the position uncertainty predicted by our EDL model\footnote{This is computed using Equations 6-8 of \cite{Boggs_2000}}, for the simulated gamma source described in Section \ref{subsec:data_sim} using a parameterization of the EDL model output described in detail in Appendix Section \ref{sec:edl_param}. The solid blue curve in Figure \ref{fig:pos_unc_effects}b plots the FWHM of a Lorentz distribution fitted to the projection of Figure \ref{fig:pos_unc_effects}a onto the $Y$-axis, with a cut on the position-related pointing uncertainty given by the $X$-axis value. The dashed green curve gives the fraction of total events remaining after applying such a cut}. With no uncertainty cut \redtwo{(i.e.~the $Y$-value of the solid blue curve at the right edge of Figure \ref{fig:pos_unc_effects}b)}, the FWHM of this distribution is \redtwo{2.788} degrees, very close to the \redtwo{2.746} degrees for the deterministic model (Figure \ref{fig:det_ARM}). \redtwo{We intend to study the extent to which a cut on this parameter can be used in concert with other standard acceptance criteria (requiring sufficiently large separation between Compton scatters, choosing low-angle Compton scatters, etc.)~to improve the overall pointing performance of the instrument.}

\section{Conclusions}
The GammaTPC is a telescope concept based on a single-phase liquid argon time-projection chamber with a novel dual-scale pixel-based readout system to enable a significant improvement in sensitivity to MeV-scale gamma-rays over previous telescopes. The novel pixel-based charge readout allows for imaging of tracks of electrons scattered by Compton interactions of incident gamma-rays. The field of image recognition using deep neural networks is quite advanced, and we take advantage of that to explore several different network designs for reconstruction and uncertainty quantification of the initial position and direction of scattered electrons. We use a CNN for a deterministic reconstruction of the electron track head, and a similar CNN for a deterministic reconstruction of the initial scattering direction of the electron. We find that the deterministic model is able to locate electron-track heads with sub-millimeter precision, and is able to effectively reconstruct the initial scatter direction as well. We explore several options for uncertainty quantification of the electron-track head location reconstruction, focusing most of our attention on Evidential Deep Learning (EDL) \citep{amini2020deep}. We find that an EDL model is able to accurately estimate its uncertainty, but that this comes at some cost in the overall accuracy in track head location estimation. An accuracy matching that of the deterministic model can be achieved by sacrificing approximately 75\% of the data. However, we further find that the EDL model produces a pointing accuracy \redtwo{close to that of the deterministic model}, given by the Angular Resolution Measurement (ARM). Removing events with large uncertainty in pointing due to large uncertainties in position location can improve the ARM further. 

There are some likely avenues for improvement. We did not emphasize optimization of training hyperparameters and procedures in this analysis, so it is feasible that changes there could improve the performance of this model. A larger data set would also likely improve the training, and open up the possibility of a deeper model, which could also improve performance. We also envision refinements in both the base model and the uncertainty estimation approach. The base model at present uses 3D convolutions as feature extractors. This operation treats the data as embedded in a space with a Euclidean metric and tends to favor shorter range spatial interactions. Furthermore, most of the voxels in each sample are empty, leading to sparse data, which CNNs were not originally designed for. In this context, we intend to apply Graph Neural Networks (GNNs) as the base model, which have shown success in reconstruction tasks for neutrino physics applications \citep{ju2020graph,thais2022graph}. With respect to the uncertainty estimation, the absence of the off-diagonal terms in the predicted covariance between the different spatial directions likely led to a tendency towards larger prediction intervals than would have existed if off-diagonal terms were included. To address this issue, we intend to utilize Variational-Inference based Bayesian Neural Networks and \textit{a posteriori} Laplace's-Approximation based uncertainty estimation. \redtwo{Finally, a model that can predict both the Compton-scattered electron's initial direction and an uncertainty estimate would enable us to perform a similar operation, placing a threshold on that uncertainty estimate. Given that, as we showed in Section \ref{subsec:direction_model}, higher-energy tracks tend to have \textit{superior} direction reconstruction (as opposed to the opposite effect with position reconstruction), combining these two estimates could be powerful. We intend to develop an uncertainty-estimating model for the direction reconstruction going forward.}

\appendix
\restartappendixnumbering

\section{Parameterization of model outputs for use with \texttt{revan}}
\label{sec:appendix}

\subsection{Deterministic track head location only model}
\label{sec:det_param}
To simulate the model response on the MEGAlib data set of Compton scatters, we interpolate in electron energy between the fitted standard deviations in Figure \ref{fig:fitted_error} to get an effective \redtwo{Gaussian} error distribution as a function of initial electron energy, and sample from those distributions for each energy deposition. We then apply this sampled reconstruction error to all 3 dimensions (each one independently sampled) for a given Compton scatter to simulate the reconstruction error that would arise from using this model. We do not simulate incident gamma-rays with energies above 1,000 keV, therefore there are never electrons with energies higher than the highest value in our training data set, which is also 1,000 keV. Compton scattered electrons with energies below 50 keV are treated with the same error distribution as the 50 keV electron data, as at those energies the electron track is point-like on the scale of the pixel readout. \redtwo{For simplicity, and since the final detector configuration not yet determined (including the full drift length), we use the 5 cm drift model for all scatters. Regardless, the RMS error of the model predictions does not change much between 5 and 10 cm drift depth (see Figure \ref{fig:deterministic_5_cm_drift_abs_error}), and furthermore this part of the analysis is only relevant for Figure \ref{fig:det_ARM}.}

\subsection{Evidential Deep Learning model}
\label{sec:edl_param}
\begin{figure}
    \centering
    \includegraphics[width=\columnwidth, trim={0.2cm 0.7cm 0.6cm 0.0cm},clip]{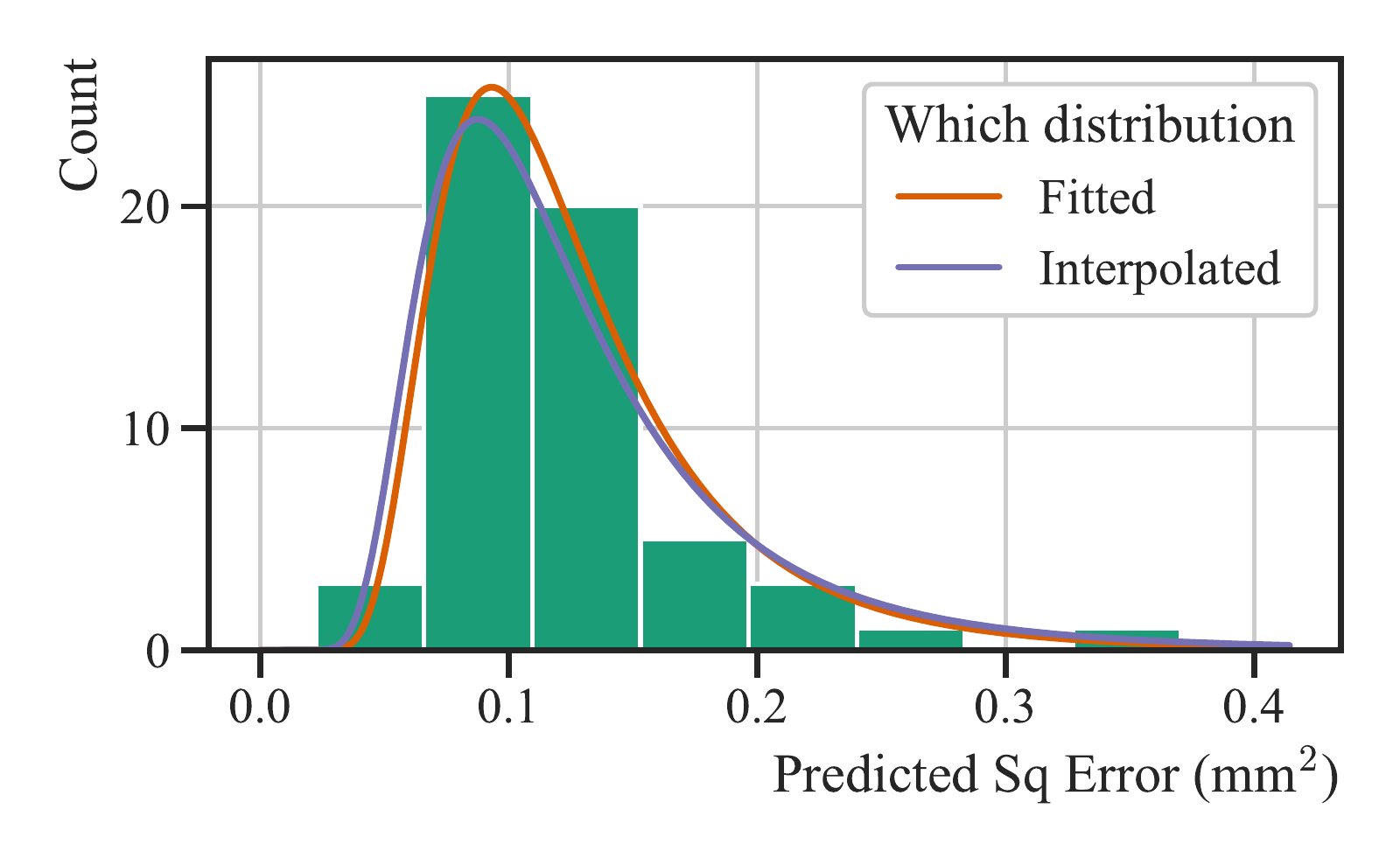}
    \caption{The distribution of the predicted squared error for 1,000 keV gamma-rays at a true \redtwo{squared} error of \redtwo{0.09} mm${}^2$. The orange line is a fit to an inverse-gamma distribution, while the purple line is the inverse-gamma distribution predicted by the parameterized model.}
    \label{fig:inv_gamma_dist}
\end{figure}

\begin{figure}
    \centering
    \includegraphics[width=\columnwidth, trim={0.7cm 0.7cm 0.6cm 0.65cm},clip]{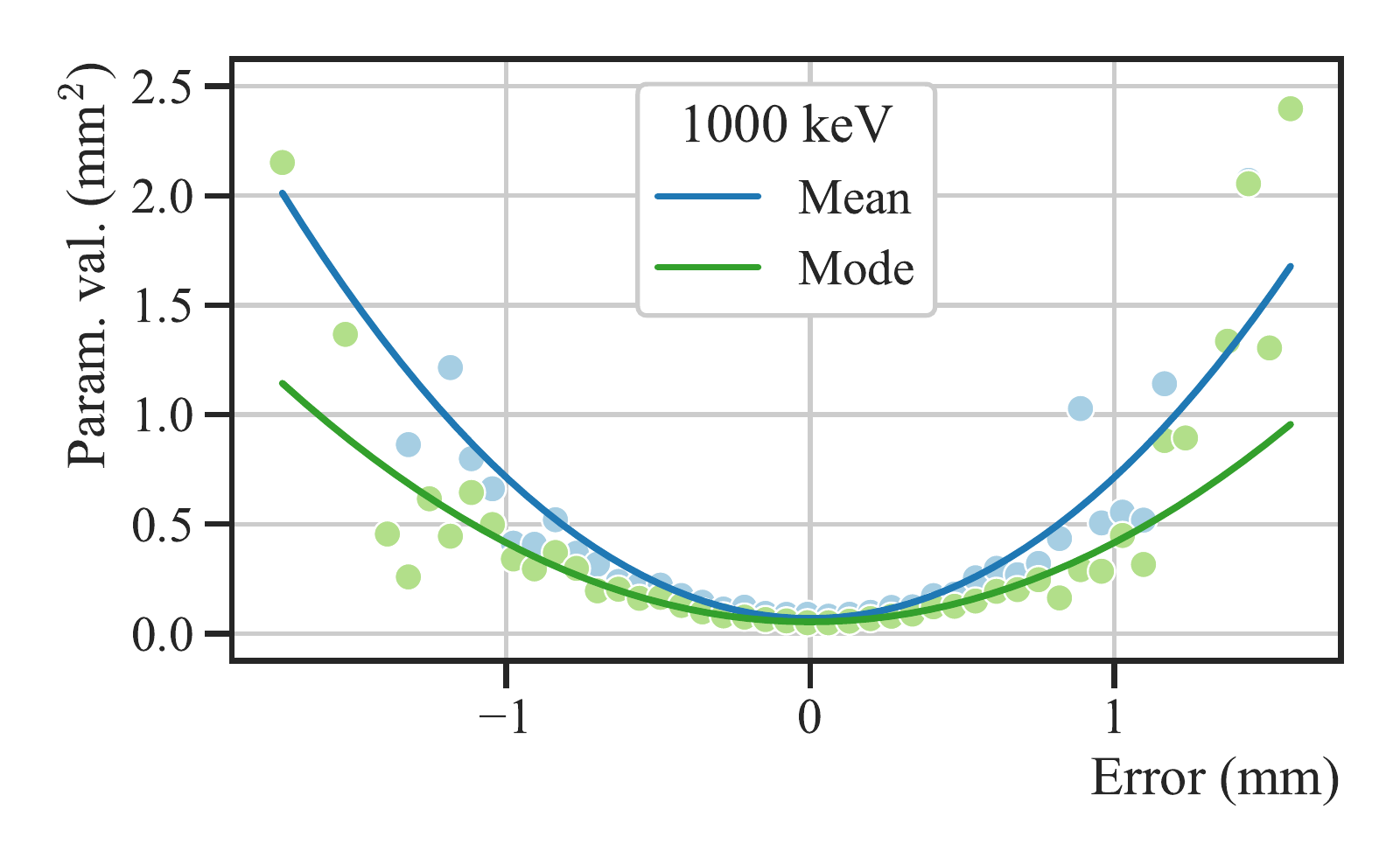}
    \caption{Each point represents the mean (blue) or mode (green) for an inverse-gamma distribution fitted to a slice of the distribution shown in the top panel of Figure \ref{fig:unc_vs_error}. The means/modes themselves follow a quadratic relationship with the true error.}
    \label{fig:inv_gamma_params}
\end{figure}

\redtwo{As we established in Section \ref{subsec:edl_model}, the predicted squared error and the true error follow a parabolic relationship. Figure \ref{fig:inv_gamma_dist} presents a slice of the 2D distribution from the top panel of Figure \ref{fig:unc_vs_error} at a constant true error of 0.3 cm (or a constant true squared error of 0.09 cm${}^2$), showing that the predicted squared error is indeed inverse-gamma distributed.} The means and modes of these inverse-gamma distributions at each bin in true error can be reasonably approximated by parabolas centered at $X=0$, seen in Figure \ref{fig:inv_gamma_params}. We can then use these parabolas to obtain shape parameters for the appropriate inverse-gamma distribution values for a given true error.

\begin{figure}
    \centering
    \includegraphics[width=\columnwidth, trim={0.6cm 0.7cm 0.6cm 0cm},clip]{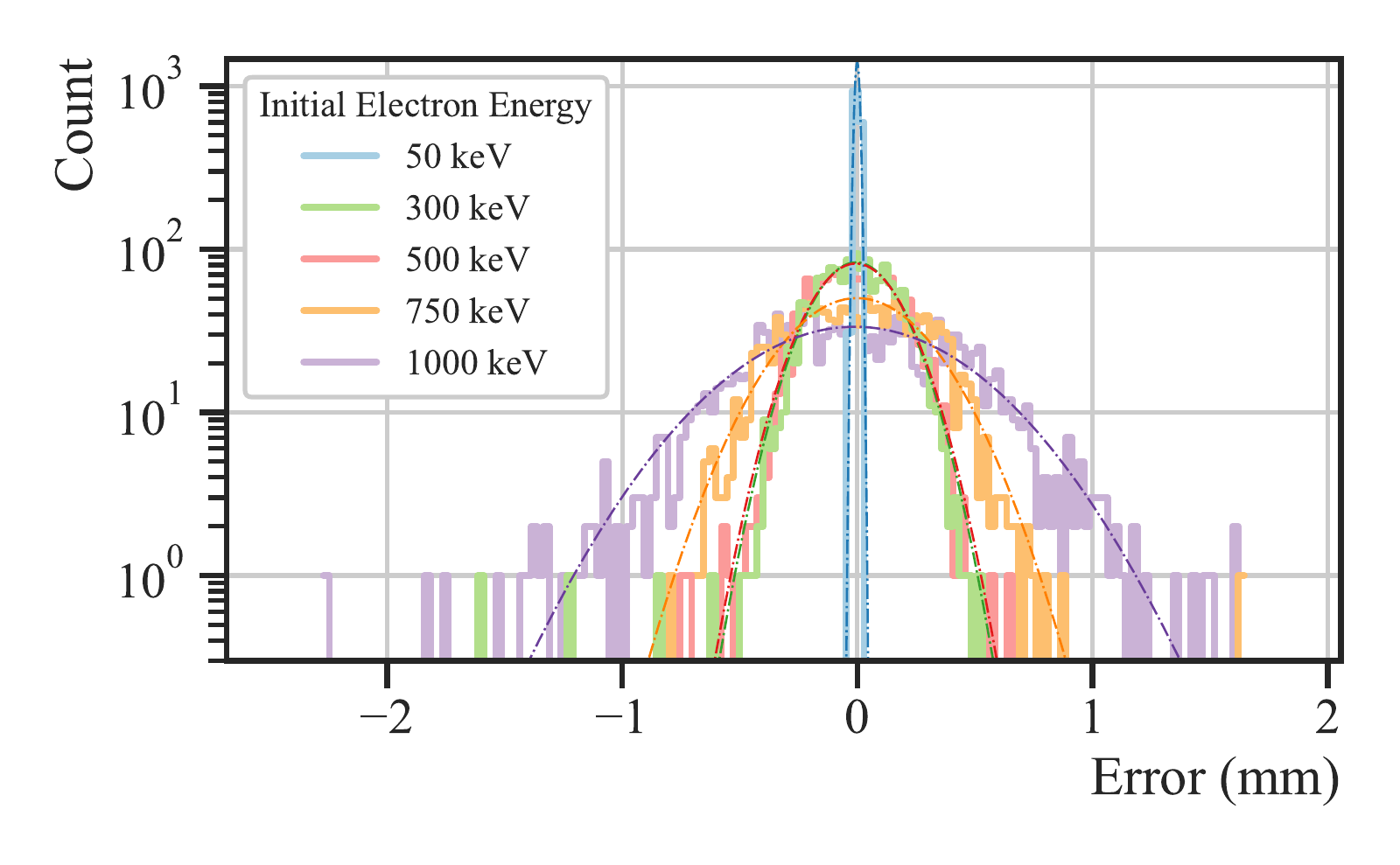}
    \caption{Distributions of true errors at the simulated energies. Predictions for all 3 cardinal directions are included in these distributions, because they were treated independently by the neural network.}
    \label{fig:edl_fitted_mean_err}
\end{figure}

The steps for sampling a pair of true error and predicted uncertainty for a Compton scatter at a given energy are then as follows: First, we obtain the standard deviation of the Gaussian distribution describing the true error for a given energy by interpolating between the true error distributions fitted at the training energies. These distributions are shown in Figure \ref{fig:edl_fitted_mean_err}. We again round up to 50 keV any scatters with an energy less than 50 keV. We next use our sampled error to draw a predicted uncertainty from an inverse-gamma distribution. To do this, we interpolate in energy parameters for parabolas characterizing the mean and mode of the inverse-gamma distribution based on the values fitted at the training energies. We then use the two parabolas to get the mean and mode of our inverse-gamma distribution at the value of our true error sample. We then draw a random value from the inverse-gamma distribution for the predicted squared error (e.g.~Figure \ref{fig:inv_gamma_dist}), and take its square root to obtain the predicted uncertainty. \redtwo{We again use the results of the 5 cm drift model for all scatters for the same reasons stated in Section \ref{sec:edl_param}. This parameterization is used only to produce Figure \ref{fig:pos_unc_effects}.}

\bibliography{gammatpc}{}
\bibliographystyle{aasjournal}



\end{document}